\documentclass[aps,pra,twocolumn,groupedaddress,showkeys,longbibliography,nofootinbib]{revtex4-1}

\usepackage{graphicx,amsmath,amssymb}
\usepackage{times,color}

\newcommand{\bra}{\langle}
\newcommand{\ket}{\rangle}
\newcommand{\Ham}{H}
\newcommand{\eps}{\varepsilon}
\newcommand{\lam}{\lambda}
\newcommand{\re}{\operatorname{\mathrm{Re}}}
\newcommand{\im}{\operatorname{\mathrm{Im}}}

\begin{document}

\title{The arrow of time in open quantum systems and dynamical breaking of the resonance-antiresonance symmetry}

\author{Gonzalo Ordonez}
\affiliation{Department of Physics and Astronomy, Butler University,
4600 Sunset Ave, Indianapolis, Indiana 46208, USA}
\email[]{gordonez@butler.edu}

\author{Naomichi Hatano}
\affiliation{Institute of Industrial Science, 
The University of Tokyo, 
4-6-1 Komaba, Meguro-ku, Tokyo, 153-8505, Japan}
\email[]{hatano@iis.u-tokyo.ac.jp}

\date{\today}

\begin{abstract}
Open quantum systems are often represented by non-Hermitian effective Hamiltonians that have complex eigenvalues associated with resonances. In previous work we showed that the evolution of tight-binding open systems can be represented by an explicitly time-reversal symmetric expansion involving all the discrete eigenstates of the effective Hamiltonian. These eigenstates include complex-conjugate pairs of resonant and anti-resonant states. An initially time-reversal-symmetric state contains equal contributions from the resonant and anti-resonant states. Here we show that as the state evolves in time, the symmetry between the resonant and anti-resonant states is automatically broken, with resonant states becoming dominant for $t>0$ and anti-resonant states becoming dominant for $t<0$. Further,  we show that there is a time-scale for this symmetry-breaking, which we associate with the ``Zeno time.''  We also compare the time-reversal symmetric expansion with an asymmetric expansion used previously by several researchers. We show how the present time-reversal symmetric expansion bypasses the non-Hilbert nature of the resonant and anti-resonant states, which previously introduced exponential divergences into the asymmetric expansion.
\end{abstract}



\maketitle

\section{Introduction}
The origin of the ``the arrow of time'' in nature is a subtle problem that has occupied physicists for many years. Part of the  problem is that, except for models of  weak nuclear interactions, the equations of all microscopic physical models are reversible, or time-reversal invariant [T-invariant]. In quantum mechanics this means that the generator of motion (the Hamiltonian for Schr\"odinger's equation) commutes with the anti-linear time-reversal operator $T$. 
 In contrast, macroscopic models of phenomena such as  diffusion, quantum decoherence  or radioactive decay are described by irreversible equations, which break T-invariance because their generator of motion does not commute with $T$. The question is how to relate the reversible, microscopic equations to the irreversible, macroscopic ones.

There have been several approaches to deal with this question. A recent experimental study  \cite{Batalhao2015} has shown that a microscopic quantum system can evolve irreversibly, with irreversibility being characterized by positive entropy production. The arrow of time in this experiment is attributed to the explicit time dependence of the Hamiltonian and the specific choice of the initial state, which break the time reversal invariance of the system. In the present paper, however, we will consider a time-independent Hamiltonian and an explicitly time-reversal invariant initial condition; we will show that we can nevertheless characterize the irreversibility of the system by its degree of resonance-antiresonance symmetry breaking. 

Another approach dealing with the question of irreversibility proposes a time-asymmetric quantum mechanics, formulated in a ``rigged Hilbert space''~\cite{Bohm89, Bohm2011, delaMadrid2012}. A  related approach~\cite{Prigogine73,Tasaki91,Petrosky96,Petrosky97,Ordonez01,Petrosky01}, which we will review in more detail here, is the following: even though the equations of motion are reversible, one can isolate components of the evolving quantities that break T-invariance separately. These components obey strictly irreversible equations, which do not result from any approximations. 
In many cases, these irreversible equations closely reproduce the macroscopic irreversible equations. Therefore this approach provides a rigorous link between reversible and irreversible equations. However, here the breaking of T-invariance is introduced ``by hand;'' the separation of irreversible components is not unique. Moreover, the irreversible components do not reside in the Hilbert space. This manifests itself as an exponential growth of these components in space-time regions that are not causally connected to the initial state. 

In this paper we will present a new approach  to describe irreversibility in open quantum systems, based on Ref.~\cite{Hatano14}, in which we introduced an explicitly T-symmetric decomposition of evolving states for open quantum systems described by tight-binding models. This decomposition involves a sum over all the discrete eigenstates of the Hamiltonian, including eigenstates with complex eigenvalues. The sum is T-symmetric because complex-eigenvalue states appear in  complex conjugate pairs, corresponding to resonances and anti-resonances. 

We remark that a similar decomposition applied to models in continuous space instead of the discrete space of the tight-binding models has been previously presented in Refs.~\cite{GarciaCalderon76,Tolstikhin98,Tolstikhin06,Tolstikhin08,GarciaCalderon10, GarciaCalderon12,Brown16}. An even earlier paper by More~\cite{More71} presented a discrete-state decomposition of the Green's function by invoking the Mittag-Leffler expansion.
Our approach with the use of the tight-binding models demonstrates that the infinite space of open quantum systems does not need to be uncountable infinity but can be countable infinity.

The T-symmetric decomposition gives the following description of an evolving quantum state: If at $t=0$ the state  is even with respect to time reversal, complex-conjugate resonant and anti-resonant components have equal weights in the decomposition. However,  for $t\ne 0$, the weights change. For example for $t>0$, after a time scale we will discuss, the anti-resonance component of a pair becomes negligible compared to the resonance component. The resonance-antiresonance symmetry is broken. The time evolution of the complex-conjugate pair then closely  matches the irreversible time evolution of the resonant component alone. Thus, in this new approach, like in the previous  approach described in the third paragraph, the time evolution is separated into components that break T-invariance separately. The critical difference is that in the present new approach the  time evolution automatically selects irreversible components; we do not have to select them by hand. Moreover, all components are free from unbounded exponential growth in time or space (see also Ref.~\cite{GarciaCalderon10}, where an expansion free of unbounded exponential growth in space was obtained for a continuous-space model).

Note that in  our previous paper \cite{Hatano14} we already considered the survival probability of an excited state prepared at $t=0$ and we showed that the time evolution is dominated by the resonant components for $t>0$ and by anti-resonant components for $t<0$. However, in that paper we did not discuss how this transition occurs dynamically.  This is the main new point of the present paper.  Other new results are summarized in Sec. \ref{sec:conc}.

The present paper is organized as follows. In Section~\ref{sec:Tinv} we review the concepts of T-invariance that we will discuss. In Section~\ref{sec:model} we introduce the model that we will use as an example of a system with irreversible behavior.  The system consists of an impurity coupled to an infinite wire (discrete lattice) with a single electron. We calculate the survival probability that the electron, when placed at the impurity at $t=0$, stays there for $t\ne 0$. In Sections~\ref{sec:surv} and~\ref{sec:breaking}   we formulate the survival amplitude and review the approach in which non-Hilbert eigenstates of the Hamiltonian are singled out to obtain irreversibility. In Section~\ref {sec:QEP} we summarize the T-symmetric expansion obtained in Ref.~\cite{Hatano14}, and apply it to calculate the survival probability in our model.  In Section~\ref{sec:irrev} we show that the resonance-antiresonance symmetry of the initial state is broken by time evolution, and associate this with irreversibility. In Section~\ref{sec:tscale} we estimate the time scale for the breaking of the resonace-antiresonance symmetry. In Section~\ref{sec:FM} we show that our formulation can also be applied to a model with continuous space, namely the Friedrics model, and in Section~\ref{sec:conc} we present some concluding remarks. The details of some calculations are presented in several Appendices. 
\section{Time-reversal invariance}
\label{sec:Tinv}

We start with the time-reversal operator $T$, which commutes with $H$ and is an anti-linear operator. An initial state $|\psi(0)\ket$ evolves as $|\psi(t)\ket = e^{-iH t}|\psi(0)\ket$. Therefore we have
\begin{align} \label{Trev}
T |\psi(t)\ket  = T e^{-iH t}|\psi(0)\ket =  e^{iH t} T|\psi(0)\ket.
\end{align}
Assuming $T^2=1$, we have
\begin{align} \label{Trev2}
 |\psi(t)\ket  = T e^{iH t} T|\psi(0)\ket.
\end{align}
This equation expresses time-reversal invariance. It means that a state that evolves forward in time (for $t>0$) can be obtained by first time-reversing the initial state, next evolving it backwards in time, and finally reversing it again. 

Moreover, let us assume that the state at $t=0$ is even with respect to time reversal: 
\begin{align} \label{Tinv}
 T|\psi(0)\ket = |\psi(0)\ket.
\end{align}
Due to the anti-linearity of $T$ we  have
\begin{align} \label{Tbra}
\bra\psi(0)| T|\phi\ket = \bra\psi(0)|\phi\ket^*.
\end{align} 
for any $|\phi\ket$. Then, from Eq.~\eqref{Trev}, we obtain
\begin{align} \label{Tinv2}
 T|\psi(t)\ket = |\psi(-t)\ket
\end{align}
and hence
\begin{align} \label{Tinv3}
 \bra\psi(0)|\psi(t)\ket = \bra\psi(0)|\psi(-t)\ket^*,
\end{align}
which implies
\begin{align} \label{Tinv5}
 \left|\bra\psi(0)|\psi(t)\ket\right|^2 =\left| \bra\psi(0)|\psi(-t)\ket\right|^2.
\end{align}
In other words, for a system that is T-invariant, the survival probability of a state that is even with respect to time inversion must be an even function of time. 

A recent experiment by Foroozani \textit{et al}.~\cite{Foroozani16} that monitored a quantum system continuously indeed demonstrated this fact.
By selecting  quantum trajectories that were consistent with a {\it final} (terminal) condition $C$, they assembled an exponentially growing Rabi oscillation signal (Fig.~2 of Ref.~\cite{Foroozani16}). This is the time-reversed curve of an exponentially decaying Rabi oscillation, which follows after the condition $C$ was prepared as an initial condition (Fig.~1 of Ref.~\cite{Foroozani16}).

\section{ A simple open quantum system} 
\label{sec:model}
\begin{figure}
\includegraphics[width=0.85\columnwidth]{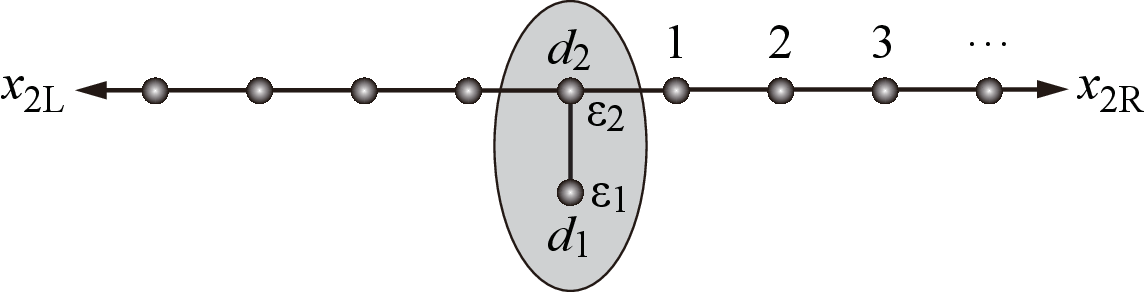}
\caption{\label{fig:Tdot} The T-shaped quantum dot model. The gray area represents the quantum dot.}
\end{figure}

We will consider a  tight-binding  model consisting of a quantum dot connected to two semi-infinite leads; see Fig.~\ref{fig:Tdot}. A single electron can move throughout the system,  hopping from site to site.

A quantity  associated with irreversible behavior is the  ``survival'' probability, which is the probability that the electron, when placed at a specific site at $t=0$, remains there for $t\ne 0$. Hereafter we will choose this site as  $d_1$. In this case, for some parameters of the system, the survival probability decays almost exponentially for $t>0$ as $t$ increases. This introduces an apparent distinction between past and future, namely the arrow of time, because exponential growth (the opposite of decay) is not observed unless very special initial conditions are chosen. 

In the remainder of this section we will present the details of our model. The complete Hamiltonian for the model shown in Fig.~\ref{fig:Tdot} is given by
\begin{align}\label{eqHam}
\Ham &\equiv \sum_{i=1}^2 \eps_{i} |d_i\ket\bra d_i| -g \left(
|d_1\ket\bra d_2 |+|d_2\ket\bra d_1|\right)\nonumber\\
& - b \sum_{\alpha=L}^R \sum_{x=1}^\infty \left(
|x_\alpha +1\ket\bra x_\alpha|+|x_\alpha\ket\bra x_\alpha+1|\right) \nonumber\\
& - \sum_{\alpha=L}^R t_{2\alpha} \left(|1_\alpha\ket\bra d_2|+|d_2\ket\bra 1_\alpha |\right),
\end{align}
where $|d_i\ket$ denotes the state for which the electron is at site $d_i$ in the dot, $|x_\alpha\ket$  denotes the electron being at the $x^{\rm th}$ site of $\alpha^{\rm th}$ lead (with $\alpha=L$ or $R$),  $\varepsilon_i$ are the chemical potentials at the sites $d_i$, $g$ is the coupling between the sites $d_1$ and $d_2$, $t_{2\alpha}$ is the coupling between the site $d_2$ and the $\alpha^{\rm th}$ lead, and $b$ is the inter-site coupling for both leads.  We assume that all parameters are real. We will also assume that all the states appearing in the Hamiltonian are even with respect to time-inversion; hence, we have $[H,T]=0$.

The dispersion relation on either lead is
\begin{align}\label{eq60}
E_k=-b(e^{ik}+e^{-ik})=-2b\cos k,
\end{align}
where $-\pi<k\leq\pi$ is the wave number limited to the first Brillouin zone. 
Because  the number of degrees of freedom on the leads is countable infinity, the wave number is limited to the Brillouin zone and the energy band has an upper bound in addition to the usual lower bound.

 The sites  $|d_1\ket$ and $|d_2\ket$ are mathematically analogous to excited and ground states, respectively, of a two-level atom, whereas the leads are analogous to a radiation field that can be emitted from and absorbed into the atom.

Introducing the states $|k_\alpha\ket$ such that $\bra x_{\alpha'}|k_\alpha\ket =\delta_{\alpha,\alpha'} \sqrt{2} \sin(kx)$, the Hamiltonian is written as $H=H_0+H_1$, where
\begin{align}\label{H0}
H_0&=\sum_{i=1}^2 \eps_{i} |d_i\ket\bra d_i| -g \left(
|d_1\ket\bra d_2 |+|d_2\ket\bra d_1|\right)\nonumber\\
&+ \sum_{\alpha=L}^R \int_{-\pi}^{\pi}\frac{dk}{2\pi} E_k |k_\alpha\ket\bra k_\alpha|,
\\
\label{H1}
H_1&=-\sum_{\alpha=L}^R t_{2\alpha} \int_{-\pi}^{\pi}\frac{dk}{2\pi} \sqrt{2}\sin(k)\left[|d_2\ket\bra k_\alpha| + |k_\alpha\ket\bra d_2|\right].
\end{align}
The Hamiltonian can then be diagonalized as follows~\cite{Newton60, Newton82}:
\begin{align}\label{eq220}
H=&\sum_{n\in\mbox{\scriptsize bound}}|\phi_n\ket E_n \bra\phi_n|
+\sum_{\alpha=L}^R \int_{-\pi}^\pi \frac{dk}{2\pi}|\phi_{k\alpha}\ket E_{k}\bra\phi_{k\alpha}|.
\end{align}
The first summation runs over all bound eigenstates $|\phi_n\ket$ with eigenvalues $E_n$, while $|\phi_k\ket$ denotes the continuum scattering eigenstates with eigenvalues $E_k$. The scattering eigenstates have the following well-known expression:
\begin{align}\label{phik}
|\phi_{k\alpha}\ket =  |{k_\alpha}\ket + \frac{1}{E_k-H+i0} H_1 |{k_\alpha}\ket.
\end{align}
The bound states are residues of the scattering eigenstates at the eigenvalues $E_n$.

\begin{figure}
\includegraphics[width=0.85\columnwidth]{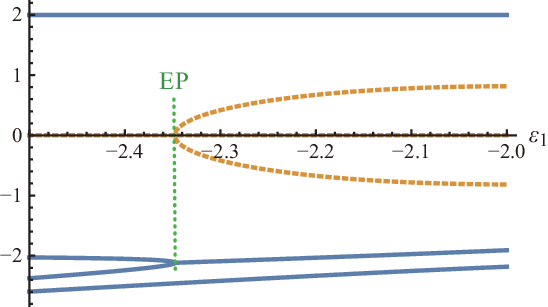}
\caption{\label{fig:EP} The real (blue solid line) and imaginary (orange broken line) parts of the discrete eigenvalues of the Hamiltonian as functions of the chemical potential $\varepsilon_1$ at site $d_1$. The imaginary part has been multiplied by $5$ for easier visualization. The vertical dotted line indicates the exceptional point (EP), where a pair of anti-bound (real) eigenvalues in the region left to the EP coalesce to form a resonance-anti-resonance pair of complex-conjugate eigenvalues in the region to the right of the EP. The bound-state eigenvalues form the top and bottom lines.
The parameters of the quantum dot are
$\varepsilon_2 = 0$, and $g=0.4$, all in units of the inter-site coupling of the leads, $b$, which we set equal to the couplings between the dot and each of the two leads: $t_{2L}=t_{2R}=b=1$.}
\end{figure}

In addition to the continuous eigenvalues $E_k$, the Hamiltonian has four discrete eigenvalues: two bound-state eigenvalues, included in Eq.~\eqref{eq220}, and either a pair of real eigenvalues corresponding to anti-bound states or a pair of complex-conjugate eigenvalues, corresponding to resonant and anti-resonant states; 
for concise reviews of resonant and anti-resonant states, see Appendix~\ref{app:review-res} of the present article as well as Section~II of Ref.~\cite{Hatano14}. 
The transition from real to complex eigenvalues occurs at  the exceptional point (EP) as the energy at site $d_1$ is increased, as exemplified in Fig.~\ref{fig:EP}.

The appearance of two complex eigenvalues makes the breaking of time-reversal symmetry possible~\cite{Garmon12}, if either one of the eigenvalues is selected to generate the time evolution. In Section~\ref{sec:irrev} we will show that such selection occurs automatically, although the system remains T-invariant.

\section{Survival amplitude}
\label{sec:surv}
\begin{figure}
\includegraphics[width=0.7\columnwidth]{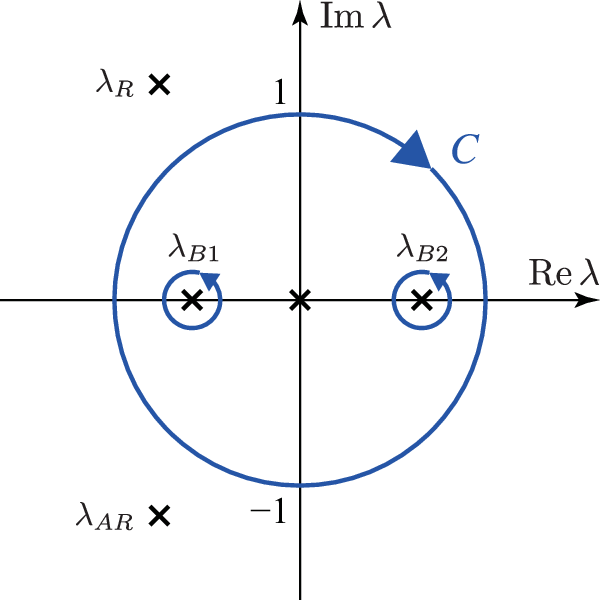}
\caption{\label{fig:Contour} The contour $C$ used in Eq.~\eqref{eq:AljLam3}. It includes the unit circle and integrations around the bound-state poles  $\lam_{B1}$ and $\lam_{B2}$. Also shown are the resonance ($\lam_R$) and anti-resonance ($\lam_{AR}$) poles as well as the pole at the origin.}
\end{figure}

The survival probability of the `excited state' $|d_1\rangle$ is given by $P(t)=|A(t)|^2$, where $A(t)$ is the survival amplitude:
\begin{align}\label{eqsurv}
&A(t) \equiv \bra d_1|e^{-iHt}|d_1\ket = \sum_{n\in\mbox{\scriptsize bound}}\bra d_1|\phi_n\ket e^{-i E_nt}  \bra\phi_n|d_1\ket \nonumber\\
&+\sum_{\alpha=L}^R \int_{-\pi}^\pi \frac{dk}{2\pi}\bra d_1|\phi_{k\alpha}\ket e^{-iE_{k}t} \bra\phi_{k\alpha}|d_1\ket,
\end{align}
which follows from Eq.~\eqref{eq220}.
As shown in Appendix~\ref{app:A}, introducing the variable $\lam=e^{ik}$ such that $E=-b(\lam+\lam^{-1})$, we can express the survival amplitude in the form
\begin{align} \label{eq:AljLam3}
& A (t)  =   \int_C \frac{d\lambda}{2\pi i \lam} \left(-\lambda+\frac{1}{\lambda}\right)e^{ib\left(\lambda+\frac{1}{\lambda}\right)t}
\frac{bg^2}{h(\lam)}\frac{1}{f(\lam)},
\end{align}
where 
\begin{align}\label{flam}
f(\lam) &= \left[-b\left(\lam+\frac{1}{\lam}\right)-\eps_{1}\right]\nonumber\\
&\times \left[-b\left(\lam+\frac{1}{\lam}\right)-\eps_{2}+\lam \sum_{\alpha} \frac{t_{2\alpha}^2}{b}\right]-g^2,
\end{align}
\begin{align}\label{hlam}
h(\lam) \equiv -b\left(\lam+\frac{1}{\lam}\right)-\eps_{1},
\end{align}
and the contour $C$ is shown in Fig.~\ref{fig:Contour}; we have assumed that the model is in such a parameter region that $f(\lam)$ has a pair of complex-conjugate roots $\lambda_R$ and $\lambda_{AR}$ corresponding to resonance and anti-resonance poles, respectively, and two real roots $\lambda_{B1}$ and $\lambda_{B2}$ corresponding to the two bound states, namely in the region to the right of the EP in Fig.~\ref{fig:EP}. As shown in Appendix~\ref{app:Eq21}, the poles of $h(\lam)$ do not contribute to the integral.

\begin{figure*}
\includegraphics[width=0.8\textwidth]{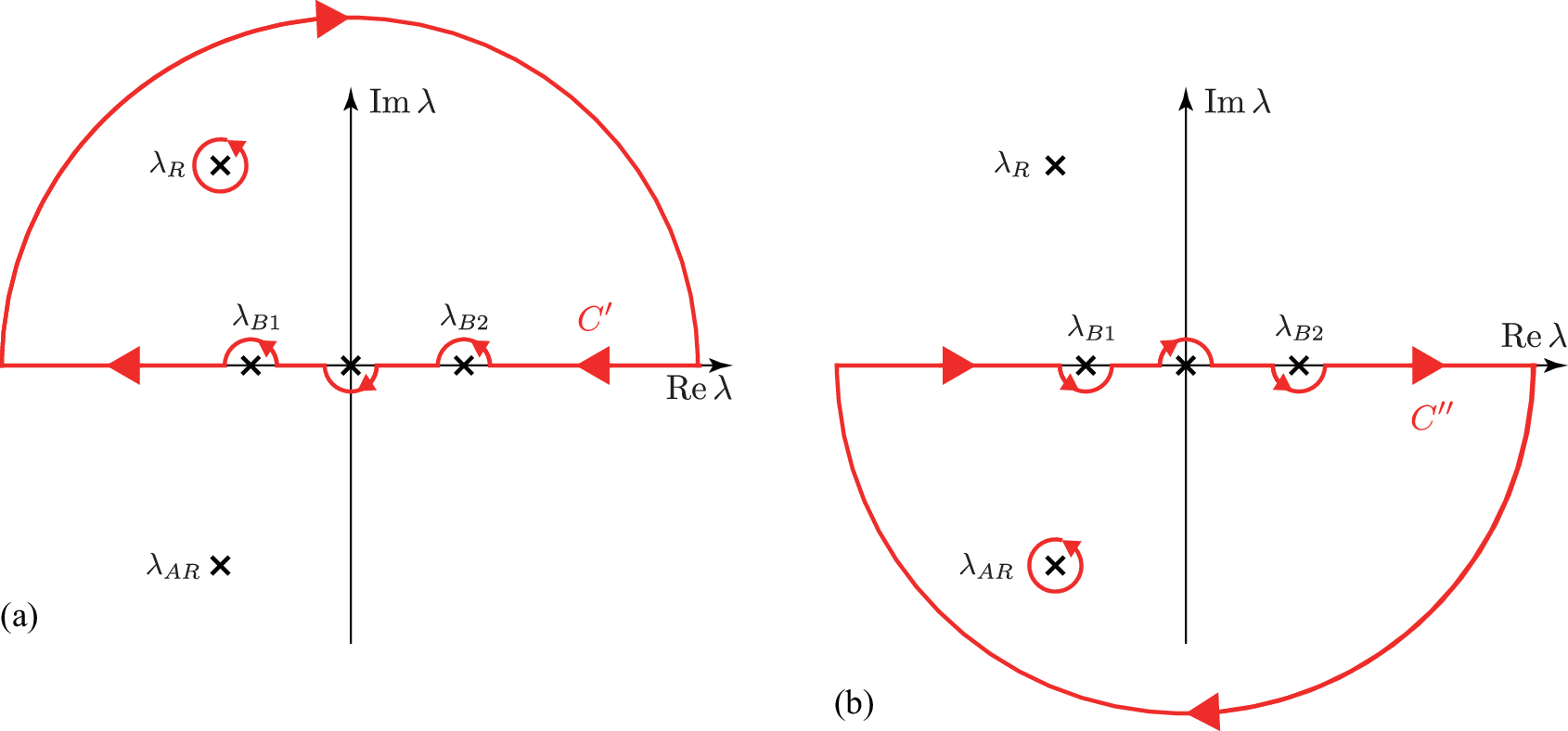}
\caption{\label{fig:ContourDef} (a) The deformed contour $C'$ used in Eq.~\eqref{eq:AljLam4} for $t>0$. It consists of the real axis going around the origin and the real poles $\lam_{B1}$ and $\lam_{B2}$ as well as the infinitely large half circle in the upper half plane. Also shown is the integration around the resonance pole $\lam_R$, which gives the first term in the right-hand-side of Eq.~\eqref{eq:AljLam4}.
(b) The deformed contour $C''$ that should be used for $t<0$.}
\end{figure*}

When $t>0$ in Eq.~\eqref{eq:AljLam3}, we can deform the contour $C$ in Fig.~\ref{fig:Contour} so that we may isolate the contribution from the residue resonant pole $\lam_R$ as shown in Fig.~\ref{fig:ContourDef}(a):
\begin{align} \label{eq:AljLam4}
& A (t)  =   \bra d_1|\phi_R\ket e^{-i E_Rt}  \bra{\tilde\phi}_R|d_1\ket \nonumber\\
&+ \int_{C'} \frac{d\lambda}{2\pi i \lam} \left(-\lambda+\frac{1}{\lambda}\right)e^{ib\left(\lambda+\frac{1}{\lambda}\right)t}
\frac{bg^2}{h(\lam)}\frac{1}{f(\lam)},
\end{align}
The first term is the residue at $\lam_R$ and is written in terms of  the right and left resonant eigenstates  of the Hamiltonian, $|\phi_R\ket$ and $\bra{\tilde\phi}_R|$ respectively, which satisfy
\begin{align} \label{Hphires}
H|\phi_R\ket = E_R |\phi_R\ket, \qquad 
\bra{\tilde\phi}_R|H = \bra{\tilde\phi}_R| E_R,
\end{align}
where $E_R = -b(\lam_R +{\lam_R}^{-1})$ is  equal to the complex resonance energy.
These states are normalized such that $\bra{\tilde\phi}_R|\phi_R\ket=1$, but they are {\it not} in the Hilbert space because the Hilbert norms $\bra{\phi}_R|\phi_R\ket$ or $\bra{\tilde\phi}_R|{\tilde \phi}_R\ket$ diverge \cite{Tasaki91}. It is worth noting that although these states grow exponentially in space representation, the normalization constant  giving $\bra{\tilde\phi}_R|\phi_R\ket=1$ can be obtained by adding a convergence factor, as shown in Appendix D of Ref.~\cite{Hatano14}.

Since the energy eigenvalue $E_R$ has a negative imaginary part, the first term in Eq.~\eqref{eq:AljLam4} represents an exponential decay for $t>0$.
The anti-resonant eigenstate (associated with the anti-resonant pole) is the complex conjugate of the resonant eigenstate as  $|\phi_{AR}\ket = |\phi_R\ket^*$, which, however, does not explicitly contribute to Eq.~\eqref{eq:AljLam4}.
The contributions from both the infinitesimal half circle around the origin and the infinite half circle in the upper half plane vanish for $t>0$ because of the exponential factor in the integrand, which is indeed the motivation for modifying the contour to $C'$ in the first place.
The integration on the real axis in Fig.~\ref{fig:ContourDef} can be evaluated by different approximations, for example the saddle-point approximation for large $t$~\cite{Hatano14} (see also Appendix \ref{app:LT}) or the short-time approximation in Section~\ref{sec:tscale}.

When $t<0$ in Eq.~\eqref{eq:AljLam3}, we are compelled to deform the contour as shown in Fig.~\ref{fig:ContourDef}(b), because contributions from both the infinitesimal half circle around the origin and the infinite half circle in the lower half plane vanish for $t<0$. After the contour deformation we obtain 
\begin{align} \label{eq:AljLam5}
& A (t)  =   \bra d_1|\phi_{AR}\ket e^{-i E_{AR}t}  \bra{\tilde\phi}_{AR}|d_1\ket \nonumber\\
&+ \int_{C''} \frac{d\lambda}{2\pi i \lam} \left(-\lambda+\frac{1}{\lambda}\right)e^{ib\left(\lambda+\frac{1}{\lambda}\right)t}
\frac{bg^2}{h(\lam)}\frac{1}{f(\lam)}.
\end{align}
Now the anti-resonant pole contributes instead of the resonant pole.
Since the energy eigenvalue $E_{AR}$ has a positive imaginary part, the first term in Eq.~\eqref{eq:AljLam5} grows exponentially  as negative $t$ increases, approaching the origin. The integration over the real axis can be evaluated by the same approximations used for the $t>0$ case.

Note that these countour deformations are the most natural choice for the evaluation of the integral in the respective cases of $t>0$ and $t<0$ because we should nullify the essential singularities at $\lambda=0$ and $|\lambda|=\infty$. 
We will show in Section~\ref{sec:QEP} by numerically evaluating the integral that the above arguments indeed give the correct behavior.

\section{Previous approach: Breaking  time-reversal invariance by hand}
\label{sec:breaking}
In this section we will summarize the previous approach~\cite{Prigogine73,Tasaki91,Petrosky96,Petrosky97,Ordonez01,Petrosky01}  for obtaining irreversible equations, using our model as a simple example.  

The main idea is to isolate terms like the first term in Eq.~\eqref{eq:AljLam4}, which break T-invariance. To do this, we  introduce the projection operators $P_1 \equiv |d_1\ket\bra d_1|$, $\Pi_R \equiv |\phi_R\ket\bra {\tilde\phi}_R|$  and ${\check \Pi}_R = 1 - \Pi_R$.  Then Eq.~\eqref{eq:AljLam4} may be written as 
\begin{align} \label{Trev33}
A(t)=\bra d_1|\xi_R(t)\ket + \bra d_1|{\check \xi}_R(t)\ket,
\end{align}
where the states
\begin{align} \label{Trev3}
 |\xi_R(t)\ket   =P_1  \Pi_R  e^{-iE_R t} |d_1\ket, \\
  |{\check \xi}_R(t)\ket   =P_1  {\check \Pi}_R  e^{-iH t} |d_1\ket
\end{align}
correspond to the first and second terms in  Eq.~\eqref{eq:AljLam4}, respectively.

The state $ |\xi_R(t)\ket $ breaks the T-invariance because by applying $T^2=1$ to it and using the antisymmetry of $T$ we obtain
\begin{align} \label{Trev5}
  |\xi_R(t)\ket  =  T e^{iE_R^* t} T P_1\Pi_R|d_1\ket \ne 
   T e^{iE_R t} T P_1\Pi_R|d_1\ket,
\end{align}
which violates  Eq.~\eqref{Trev2}.  The reason is that  the generator of motion in Eq.~\eqref{Trev3} is not  the Hamiltonian; it is $E_R$, which is  complex and no longer commutes with the $T$ operator.

 However, the complete time evolution is T-invariant, because it is generated by the Hamiltonian. This means that the second term in Eq.~\eqref{eq:AljLam4} must also break the T-invariance in such a way that the complete quantity   satisfies T-invariance. 

\begin{figure}
\includegraphics[width=0.8\columnwidth]{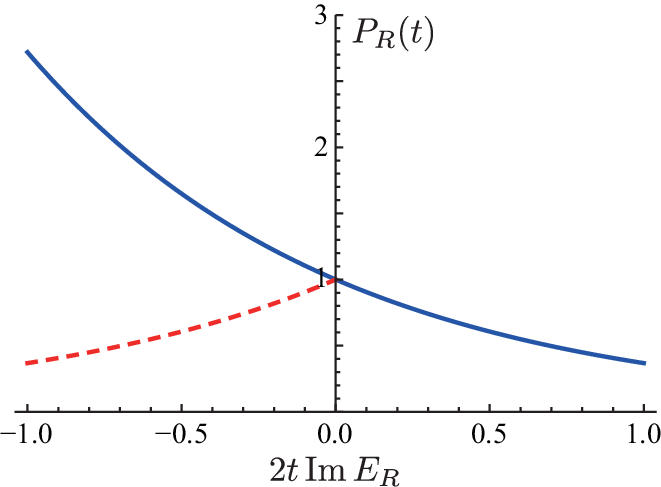}
\caption{\label{fig:ExpD} Solid line: the resonance component of the survival probability $P_R(t) = \left|\bra d_1|\xi_R(t)\ket \right|^2$. Dashed line: the anti-resonance component for $t<0$.}
\end{figure}

The state $ |\xi_R(t)\ket $  makes physical sense when $t>0$. As shown in Fig.~\ref{fig:ExpD}, it becomes unphysical for $t<0$, however, because the projected component of the survival probability then grows exponentially, unbounded, as $t$ becomes more negative. To avoid this, for $t<0$ we  have to extract the residue at the anti-resonance pole $\lam_{AR}$ instead of $\lambda_R$. Doing this will lead to exponential decay towards the past as $t$ becomes more negative. Alternatively, we can say that in this case the condition at $t=0$ is a final condition (rather than an initial condition) and for $t<0$ the survival amplitude grows exponentially as it approaches the terminal condition at $t=0$. In order for it not to grow unbounded when $t$ becomes positive, we must switch the pole that we isolate from the anti-resonant pole $\lam_{AR}$ to the resonant pole $\lambda_R$ at $t=0$.

Thus, in summary, the survival amplitude may be decomposed as a sum of terms that break T-invariance individually but added together are T-invariant as is shown in Eq.~\eqref{Trev33}.  To obtain the proper direction of the growth and the decay, we picked up a pole ``by hand.'' If the system has more than one pair of complex-conjugate eigenvalues, it  becomes somewhat arbitrary which poles we pick; the separation of irreversible components is in general not unique. Moreover, the irreversible components can become unphysical. An example of this is the $\Pi_R$ component of the initial state becoming unphysical for $t<0$.

In the following section we will present our new formulation, which relies on a T-symmetric decomposition, yet nevertheless reveals the emergence of irreversible behavior. 

\section{Present approach: The T-symmetric formulation}
\label{sec:QEP}
We will express the survival amplitude of the `excited state' $|d_1\rangle$ using the result of the quadratic-eigenvalue-problem (QEP) formalism discussed in Ref.~\cite{Hatano14}.  As shown in Ref.~\cite{Hatano14}, the survival amplitude~\eqref{eqsurv} is also expanded in the form
\begin{align} \label{eq:xrep}
 A(t) &=\sum_{n=1}^{2N}\bra d_1|\chi_n(t)\ket
\end{align}
with
\begin{align} \label{eq:Xin}
 |\chi_n(t) \ket &=\frac{1}{2\pi i}
\int_{C}d\lambda\,\left(-\lambda+\frac{1}{\lambda}\right)\exp\left[ib\left(\lambda+\frac{1}{\lambda}\right)t\right]
\nonumber\\
&\times
|\psi_n\ket\frac{\lambda_n}{\lambda-\lambda_n}\bra\tilde{\psi}_n|d_1\ket.
\end{align}
Note that $N=2$ in the present case, which is the number of sites of the dot in the gray area of Fig.~\ref{fig:Tdot}. 

The integration contour $C$ is shown in Fig.~\ref{fig:Contour}. The summation over $n$ includes resonant, anti-resonant, and bound eigenstates of the Hamiltonian. The states $|\psi_n\ket$ and $\bra\tilde{\psi}_n|$ are the corresponding right and left eigenstates of the Hamiltonian with eigenvalues $E_n=-b(\lambda_n+{\lambda_n}^{-1})$.  
They have a different normalization than the eigenstates $|\phi_n\ket$ used in Section~\ref{sec:surv} as in
\begin{align} \label{eq:phipsi}
|\phi_n\ket = (1-\lam_n^2)^{1/2}|\psi_n\ket.
\end{align}

For the present model, Eq.~\eqref{eq:xrep} can also be obtained from Eq.~\eqref{eq:AljLam3} using a partial-fraction expansion of the integrand; see Appendix~\ref{app:Eq21}.
Note, however, the essential difference between the expressions~\eqref{eqsurv} and~\eqref{eq:xrep}. The former (Eq.~\eqref{eqsurv}) is expanded in terms of the well-known complete set that consists of the bound states and the scattering states, in other words, all states on the first Riemann sheet~\cite{Newton60, Newton82}.
We expanded the latter (Eq.~\eqref{eq:xrep}) in terms of our new complete set which consists of all point spectra on the first and second Riemann sheets, namely the bound, anti-bound, resonant and anti-resonant states~\cite{Hatano14}.

Instead of separating the pole contributions in the integral over $\lam$, we will simply evaluate the integral over $\lam$ for each component of the summation in Eq.~\eqref{eq:xrep}, individually. The  eigenstate components such as  $\bra d_1|\psi_n\ket$ can be calculated using the method given  in Section VII,  Eq.~(74) of Ref.~\cite{Hatano14}.

\begin{figure}
\includegraphics[width=0.8\columnwidth]{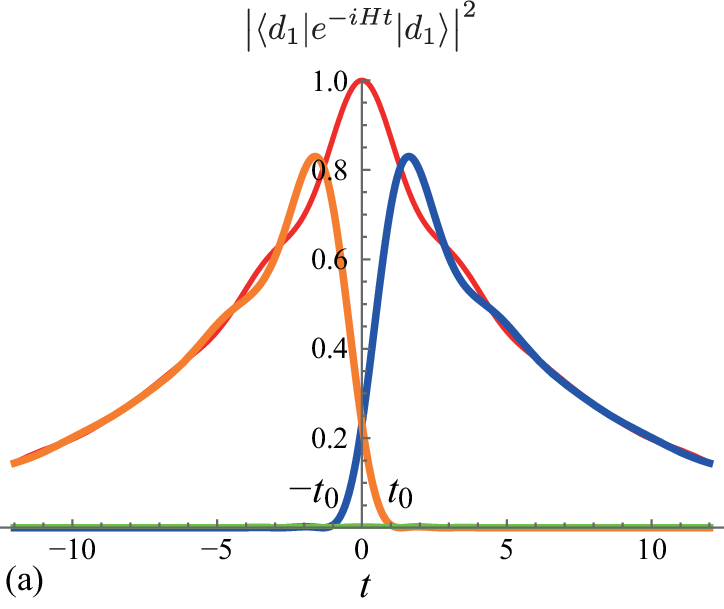}\\
\vspace{\baselineskip}
\includegraphics[width=0.8\columnwidth]{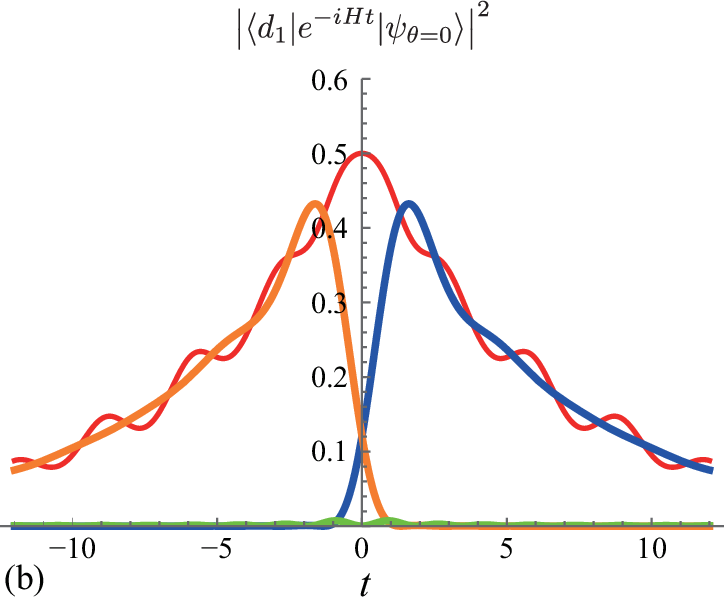}\\
\vspace{\baselineskip}
\includegraphics[width=0.8\columnwidth]{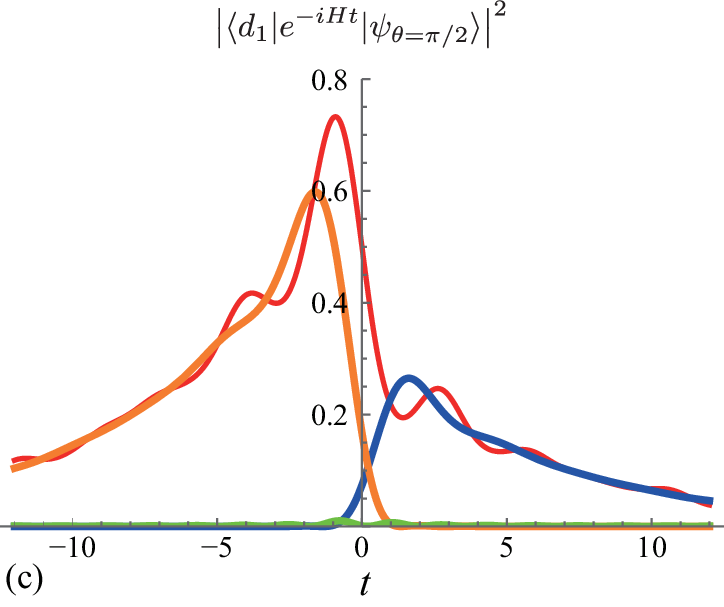}
\caption{\label{fig:SP} (a) Contributions to the total survival probability $|A(t)|^2$ of the `excited state' (red line with the highest peak) corresponding to: the resonant state $|\langle d_1|\chi_R(t)\rangle|^2$ (blue line with a lower peak on the right); the anti-resonant state $|\langle d_1|\chi_{AR}(t)\rangle|^2$ (yellow line with a lower peak on the left) and bound states $|\langle d_1|\chi_B(t)\rangle|^2$ (green line almost indistinguishable from the horizontal axis).
We here plot as the contributions the square modulus of each component, which do not add up to the plotted total survival probability.
The parameters are $b=1$, $\varepsilon_1=0.2$, $\varepsilon_2=0$, $g=0.4$, and $t_{2L}=t_{2R}=1$.
Similarly, the probability $\left|\bra d_1|\psi_\theta(t)\ket\right|^2$ with $|\psi_\theta(0)\ket$ given by Eq.~\eqref{eq:d1d2_1}, (b) for $\theta=0$ and (c) for $\theta=\pi/2$. For $\theta=\pi/2$ the resonant and anti-resonant components of the state $|\psi_\theta(0)\ket$ have different weights.}
\end{figure}

Figure~\ref{fig:SP} shows the numerically evaluated contributions $|\langle d_1|\chi_n(t)\rangle|^2$ to the survival probability corresponding to the resonant, anti-resonant and bound eigenstates. It is seen that separately, the resonant and anti-resonant contributions are not even functions of time;  we will show that they violate condition~\eqref{Tinv2} for T-invariance. However, taken as a whole the sum $\left| \langle d_1|\chi_{\rm R}(t)\rangle + \langle d_1|\chi_{\rm AR}(t)\rangle\right|^2$ adds up to a T-invariant evolution and produce an even function of time. 

\begin{figure}
\includegraphics[width=0.8\columnwidth]{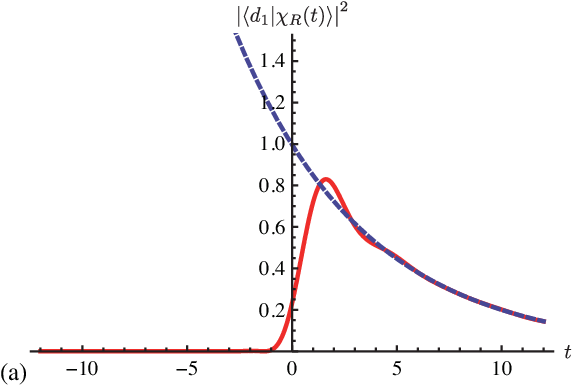}\\
\vspace{\baselineskip}
\includegraphics[width=0.8\columnwidth]{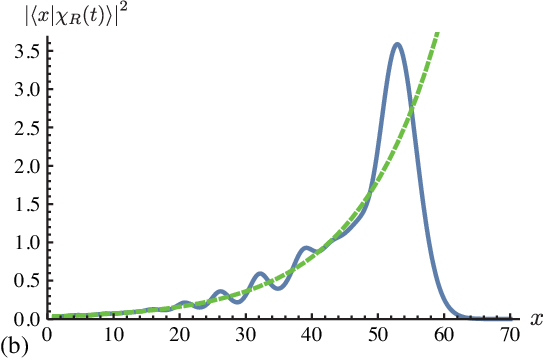}
\caption{\label{fig:NoExpBlowUp} (a) The resonant component of the survival probability, $\left|\bra d_1|\chi_R(t)\ket\right|^2$ (solid line) compared to the non-Hilbert state component $\left|\bra d_1|\phi_R(t)\ket\right|^2$ (dashed line). The parameters are the same as in Fig.~\ref{fig:SP}.
(b)   The resonant component  $\left|\bra x|\chi_R(t)\ket\right|^2$ (solid line) compared to the non-Hilbert state component $\left|\bra x|\phi_R(t)\ket\right|^2$ (dashed line) in the position representation. This figure is similar to Fig.~7.11 in Ref.~\cite{GarciaCalderon10} or Fig.~2 in Ref.~\cite{Petrosky01}, although the latter two display the complete probability density in space representation rather than the resonance component alone.}
\end{figure}

We remark that in contrast to the isolated resonant pole contribution discussed in Section~\ref{sec:breaking}, the  curve of the resonant contribution in Fig.~\ref{fig:SP} does not have an unbounded exponential growth for $t<0$; compare Fig.~\ref{fig:NoExpBlowUp}(a) with Fig.~\ref{fig:ExpD}. Similarly, the curve of the anti-resonant contribution does not grow exponentially for $t>0$. Furthermore,  the resonant or anti-resonant  components are free from any unbounded exponential growth in the position representation~\cite{GarciaCalderon10, Hatano14}, whereas the  non-Hilbertian states $|\phi_R\ket$ of $\bra{\tilde\phi}_R|$ diverge exponentially in position representation; see Fig.~\ref{fig:NoExpBlowUp}(b).
We stress that the present decomposition~\eqref{eq:xrep} produces the \textit{automatic} switching around $t=0$ from the anti-resonant contribution that grows for negative time to the resonant contribution that decays for positive time; 
we do not switch them by hand.
Another marked difference from the approach in Section~\ref{sec:breaking} is the fact that the anti-resonant contribution does not suddenly give way to the resonant contribution;
this will be the central topic of Section~\ref{sec:tscale}.

To formulate the breaking of T-invariance for the individual components of Eq.~\eqref{eq:xrep},  we consider the effect that the time-reversal operator has on each component~\eqref{eq:Xin}. 
Applying $T$ to these states takes the complex conjugate of all constants. Taking the complex conjugate of $\lam$ changes  $\lam\to\lam^{-1}$ around the unit circle, whereas $\lam$ remains unchanged on the infinitesimal circles around the bound-state poles. We can bring $\lam$ back to its initial form by making the change of variables $\lam\to\lam^{-1}$. Then we have
\begin{align} \label{eq:TXin}
T |\chi_n(t) \ket &= \frac{1}{2\pi i}
\int_{C}d\lambda\,\left(-\lambda+\frac{1}{\lambda}\right)\exp\left[-ib\left(\lambda+\frac{1}{\lambda}\right)t\right]
\nonumber\\
&\times
|\psi_n\ket^*\frac{\lambda_n^*}{\lambda-\lambda_n^*}\bra\tilde{\psi}_n|d_1\ket^*.
\end{align}
Here the resonant and anti-resonant states, $|\psi_R\ket$ and $|\psi_{AR}\ket =|\psi_R\ket^*$, are complex, while the bound-state components are real. Therefore we have
\begin{align} \label{eq:TXin2}
 & T |\chi_R(t) \ket = |\chi_{AR}(-t) \ket,\nonumber\\
 & T |\chi_{AR}(t) \ket = |\chi_{R}(-t) \ket,\nonumber\\
 & T |\chi_B(t) \ket = |\chi_{B}(-t) \ket,
\end{align}
where $B$ represents a bound state. 
We see that due to complex conjugation, the resonant and anti-resonant states  break T-invariance, while the bound states do not (compare Eq.~\eqref{eq:TXin2} with Eq.~\eqref{Tinv2}).  However, the summation of resonant and anti-resonant components satisfies T-invariance:
\begin{align} \label{eq:TXin3}
T \left(|\chi_R(t)+|\chi_{AR}(t)\ket\right)  =\left(|\chi_R(-t)+|\chi_{AR}(-t)\right).
\end{align}

So far, we have considered the decay of the `excited state,' choosing $|d_1\rangle$ as the initial state, but we can also consider an initial state of the form
\begin{align} \label{eq:d1d2_1}
|\psi_\theta(0) \ket =\frac{1}{\sqrt{2}}\left(|d_1\ket + e^{i\theta} |d_2\ket\right),
\end{align}
which is generally not  even with respect to time-inversion.
Under the time reversal we have 
\begin{align} \label{eq:d1d2_2}
 \bra d_1 | T |\psi_\theta(t) \ket= \bra d_1 | T e^{-iHt} |\psi_\theta(0) \ket  = \bra d_1 |  \psi_{-\theta}(-t) \ket 
 \end{align}
Thus, from Eq.~\eqref{Tinv3} we obtain 
\begin{align} \label{eq:d1d2_3}
\bra d_1 |  \psi_{-\theta}(-t) \ket =  \bra d_1  |\psi_\theta(t) \ket^*,
 \end{align}
which means that the probability $\left|\bra d_1  |\psi_\theta(t) \ket\right|^2$ is no longer an even function of time, unless $\theta=0$ or $\theta=\pi$. This is demonstrated in Fig.~\ref{fig:SP}~(b) and~(c). 

The components of $|\psi_\theta(t)\rangle$, 
\begin{align} \label{eq:TXith}
|\chi_n(t,\theta) \ket &= \frac{1}{2\pi i}
\int_{C}d\lambda\,\left(-\lambda+\frac{1}{\lambda}\right)\exp\left[ib\left(\lambda+\frac{1}{\lambda}\right)t\right]
\nonumber\\
&\times
|\psi_n\ket\frac{\lambda_n}{\lambda-\lambda_n}\bra\tilde{\psi}_n|\psi_\theta\ket,
\end{align}
satisfy 
\begin{align} \label{eq:TXith2}
\bra d_1|\chi_{\rm R}(t,\theta) \ket  = \bra d_1|\chi_{\rm AR}(-t,-\theta) \ket^*
\end{align}
and
\begin{align} \label{eq:TXith4}
\bra d_1|\chi_{\rm B}(t,\theta) \ket  = \bra d_1|\chi_{\rm B}(-t,-\theta) \ket^*,
\end{align}
which shows that the bound-state components and the sum of resonance and anti-resonance components of the probability are even functions of time only if $\theta=0$ or $\theta=\pi$. 

\section{Automatic breaking of resonance-antiresonance symmetry}
\label{sec:irrev}
Despite  the symmetry of the survival probability around $t=0$  in Fig.~\ref{fig:SP}(a), we can still say that the exponential decay process has an irreversible nature; starting at the dot $d_1$ at $t=0$, for $t>0$ the electron enters the infinite wire and never comes back. (The same could be phrased for $t<0$ but in this context we usually do not say that an electron moves backward in time.) 

In the following, we will argue that this irreversible nature emerges as a  breaking
of the resonance-antiresonance symmetry that exists in the initial state, which we assume is   invariant under  time inversion, namely $T|\psi(0)\ket =|\psi(0)\ket$. As a result of this invariance,  the resonant and anti-resonant components of the survival amplitude have equal magnitudes at $t=0$. For example, when the initial state is $|d_1\ket$ we have $\bra d_1|\chi_R(0)\ket = \bra d_1|\chi_{AR}(0)\ket^*$.  The initial state thus has a resonance-antiresonance symmetry.

However, this symmetry is automatically broken by time evolution, as seen in Fig.~\ref{fig:SP}(a). The figure shows that for $t>0$ the anti-resonant component is  significantly suppressed after a time $t_0$.  Figure~\ref{fig:SP}(a) also shows  that, as the anti-resonance component becomes negligible for $t>0$, the whole survival probability (red line with the highest peak) approaches the contribution due to the resonant component  alone (blue line with a lower peak on the right), which breaks T-invariance by itself. In this way the complete time evolution mimics the time evolution of the strictly irreversible resonant component. 

Note that  in the example that we are considering there are no other resonance/anti-resonance pairs  and, moreover, the bound states give a negligible contribution.  For more complicated systems, there can be more than one resonance/anti-resonance pair; the breaking of resonance-antiresonance symmetry would then manifest itself within each pair, as the time evolution alone would suppress one component of the pair relative to the other component. The bound-state contributions (even if they are not negligible) are only trivially affected by the $T$ operator, since they satisfy T-invariance by themselves, i.e., $\bra d_1|\chi_{\rm B}(t) \ket  = \bra d_1|\chi_{\rm B}(-t,) \ket^*$. 

To quantify the extent  of resonance-antiresonance symmetry breaking,  we consider  the ratio of the resonant-state contribution  to the anti-resonant state contribution within the pair:
\begin{align} \label{ratio1}
r=\left|\frac{\bra d_1|\chi_R(t)\ket}{\bra d_1|\chi_{AR}(t)\ket}\right|^2 = \left|\frac{\bra d_1|\chi_R(t)\ket}{\bra d_1|\chi_{R}(-t)\ket}\right|^2.
\end{align}
If  the resonance-antiresonance symmetry were completely broken, then $r$ would be  $\infty$ for $t>0$ and $0$ for  $t<0$. 

\begin{figure}
\includegraphics[width=0.9\columnwidth]{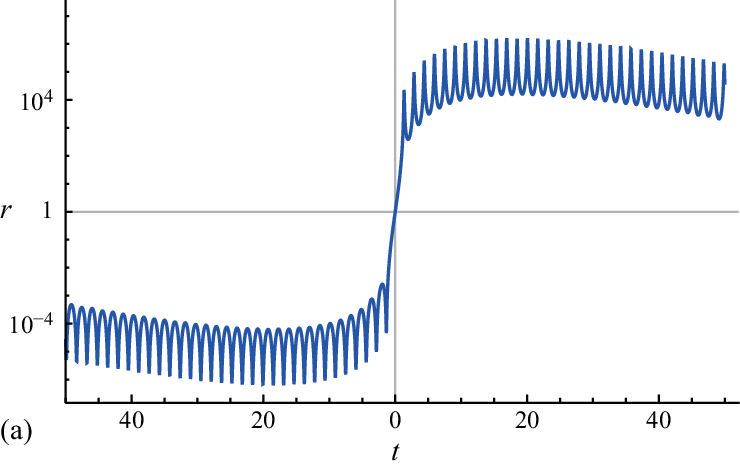}\\
\vspace{\baselineskip}
\includegraphics[width=0.9\columnwidth]{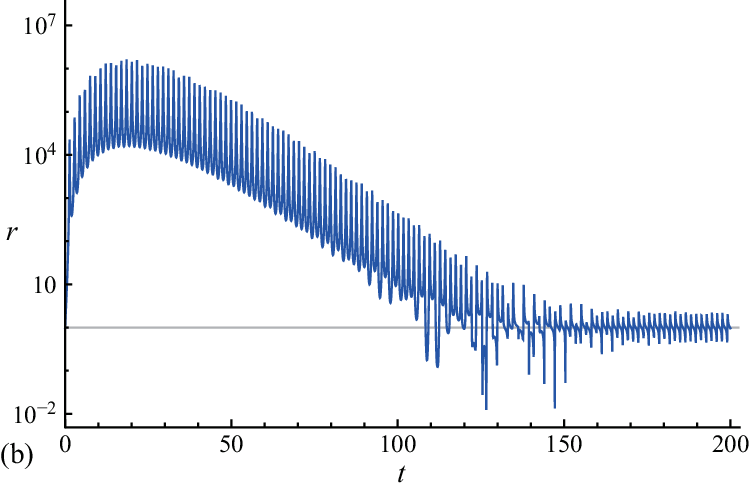}\\
\vspace{\baselineskip}
\includegraphics[width=0.9\columnwidth]{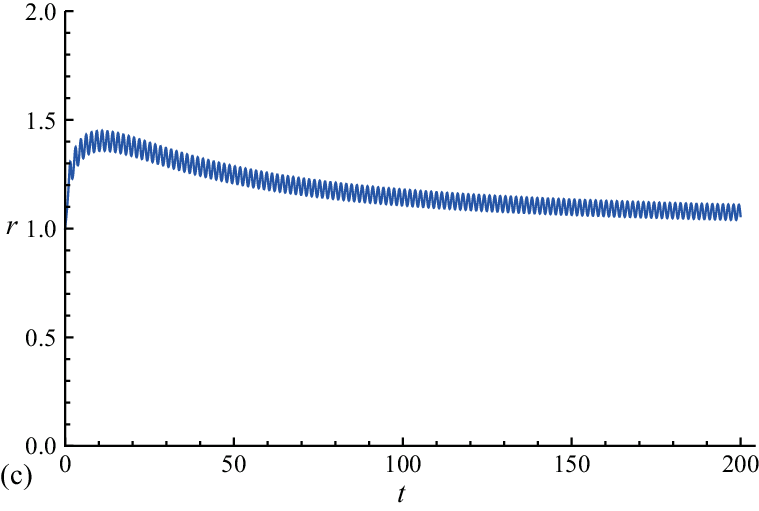}
\caption{\label{fig:Ratio} (a) The ratio $r$ of the resonant component of the survival probability to the anti-resonant component in a semi-logarithmic scale. The parameters are the same as in Fig.~\ref{fig:SP}.
(b) The same in a longer time scale.
(c) The ratio with $\varepsilon_1$ changed to $-2.347528$, when the pair of resonant and anti-resonant poles are near an exceptional point. The rapid oscillations seen in all the figures have a period inversely proportional to the band-edge energies with $|E|=2$ (see Eq.~\eqref{eq:fn5}). }
\end{figure}

Note that at $t=0$ we have $r=1$; this reflects the invariance of the state $|d_1\ket$ with respect to time reversal. 
However, as $t$ increases the ratio quickly jumps from $r=1$ at $t=0$ to a large value, as shown in Fig.~\ref{fig:Ratio}(a), meaning that the resonant-state contribution becomes much larger than the anti-resonant contribution.
This jump in $r$ occurs around the time $t=t_0\simeq 1$, which we will relate to the system parameters in Section~\ref{sec:tscale}.
Conversely,  the anti-resonant contribution dominates for $t<-t_0$. For both positive or negative times, the larger $\left|\log r\right|$ is, the more the time evolution approaches a complete breaking of resonance-antiresonance symmetry.
We stress again that since the time evolution alone suppresses the anti-resonances for $t>t_0$ (and the resonances for $t<-t_0$), we do not have to isolate the resonance or anti-resonance contributions by hand. This happens automatically. 

We remark that for very long time the regime of exponential decay is known to be replaced by the ``long-time regime" of an inverse power-law decay~\cite{Garmon13}.
Indeed, the ratio~\eqref{ratio1} settles back to unity as shown in Fig.~\ref{fig:Ratio}(b), which indicates that the irreversible behavior disappears at very long-time scales. This is proved in Appendix~\ref{app:LT}.

Another remark is that if the chemical potentials $\varepsilon_i$ of the impurity are adjusted, the pair of the resonant and anti-resonant eigenvalues (in the region right of the EP in Fig.~\ref{fig:EP}) can approach the exceptional point (EP), where these eigenvalues converge into a single real energy and subsequently split into two anti-bound real eigenvalues~\cite{Garmon12} (in the region left of the EP in Fig.~\ref{fig:EP}), which satisfy T-invariance individually. As shown in Fig.~\ref{fig:Ratio}(c), when the resonance and anti-resonance pair is near the EP, the change of the ratio $r$ is much less dramatic than in Fig.~\ref{fig:Ratio}(a). The time evolution then hardly has irreversible nature, because both the resonant and anti-resonant components have significant weight throughout the whole time evolution.

When there are no resonant and anti-resonant pairs, only bound and anti-bound states (i.e., to the left of the EP in Fig.~\ref{fig:EP}), then there is no breaking of resonance-antiresonance symmetry to speak of; the system has no irreversible behavior then.


\section{Time scale for the  breaking of resonance-antiresonance symmetry}
\label{sec:tscale}
In Fig.~\ref{fig:SP}(a), it is seen that the resonant-state contribution to the survival amplitude
does not  jump discontinuously  at $t=0$; instead it is a continuous function of time:  it increases gradually starting at a negative time $-t_0$, then it reaches a peak after $t=0$ and eventually decays exponentially. Similarly, the antiresonance-state contribution does not suddendly disappear for positive times; instead, it leaks into the postive-time region and becomes negligible after the positive time $t_0$. The time range $[-t_0,t_0]$ is a time region during which the resonance-antiresonance is unbroken; outside of this range the symmetry is broken.  In the following, we will estimate the magnitude of the time $t_0$, which will turn out to be close to the Zeno time (note that our present model does not include monitoring; the connection with the Zeno time will be done based on the conditions that the unmeasured survivival probability must meet for the Zeno effect to occur).

In Appendix~\ref{app:B} we derive the following analytic expression for the resonant component of the survival amplitude:
\begin{align} \label{eq:Acut2}
  &A_{{\rm R}}(t) =\bra d_1|\chi_R(t)\ket
=
 \bra d_1|\psi_R\ket \bra\tilde{\psi}_R| d_1 \ket \nonumber\\
&\times e^{-iE_Rt}\left[ 1
 -i  \lambda_R\int_{0} ^t dt' \, e^{i E_R t'}\frac{J_1(2bt')}{t'} \right],
\end{align}
which can also be written as 
\begin{align} \label{eq:Alongt}
  &A_{{\rm R}}(t)
=
\bra d_1|\phi_R\ket \bra\tilde{\phi}_R| d_1 \ket \nonumber\\
&\times e^{-iE_Rt}\left[ 1
 + \frac{i  \lambda_R}{1-\lambda_R^2} \int_{t} ^\infty dt' \, e^{i E_R t'}\frac{J_1(2bt')}{t'} \right];
\end{align}
see Appendix~\ref{app:C}.
The former expression~\eqref{eq:Acut2} is useful to calculate the short-time evolution, whereas the latter~\eqref{eq:Alongt} is useful for long times.  We remark that the first term in brackets in Eq.~\eqref{eq:Alongt} gives the resonant eigenstate projection $\Pi_R$ discussed in Section~\ref{sec:breaking}. 

Hereafter we will only use Eq.~\eqref{eq:Acut2}. We will  estimate the time it takes for the time-evolved state  $|d_1\ket$ to approach the irreversible evolution due to the resonant state alone.
We start with the following short-time approximation: $J_1(2bt) \approx bt$.   The resonant component of the survival probability  is then given by
\begin{align} \label{eq:rescontrib2}
& P_R(t) =\left|A_R(t)\right|^2\nonumber\\
&\approx  \left|\bra d_1|\psi_{\rm R}\ket\bra\tilde{\psi}_{\rm R}|d_1\ket e^{-iE_{\rm R}t}\left[ 1
 -\frac{b\lambda_{\rm R}}{E_{\rm R}} \left(e^{iE_{\rm R}t}-1\right)\right]\right|^2.
\end{align}

\begin{figure}
\includegraphics[width=0.85\columnwidth]{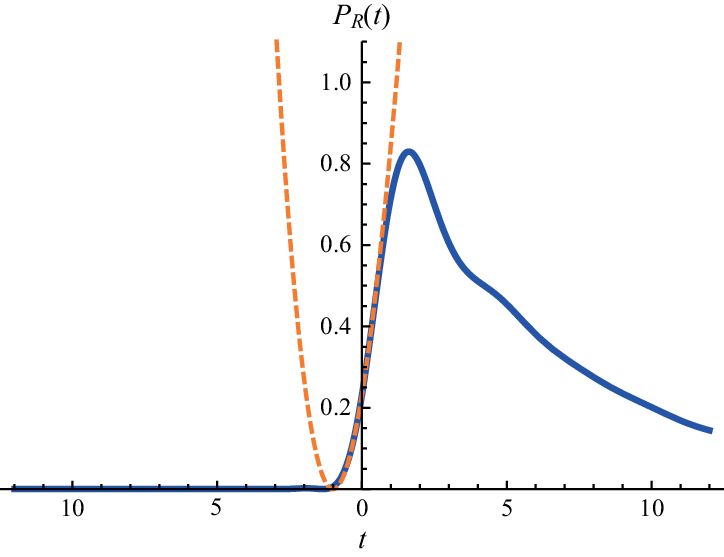}
\caption{\label{fig:surv_res} Solid line: exact contribution of the resonant state to the survival probability. Dashed line: approximation valid for short time. Parameters are $b=1, \varepsilon_1=0.2, \varepsilon_2=0, g=0.4, t_{2L}=t_{2R}=1$. For these parameters we have $E_{\rm R}=0.199675 - 0.0803343 i$, $\lambda_{\rm R}=-0.103865 + 1.03599\, i$ and $|\lambda_{\rm R}| = 1.04118$. }
\end{figure}
Figure~\ref{fig:surv_res} shows a numerical plot of the resonant-state contribution to the survival probability as a function of time.  The solid line is the  exact  expression, while the dashed line shows the short-time approximation, Eq.~\eqref{eq:rescontrib2}.  It is seen that the short-time approximation is close to zero for a negative time $-t_0$. For times earlier than $-t_0$, the exact probability $P_{\rm R}(t)$ is approximately zero and the anti-resonance contribution  dominates (see Fig.~\ref{fig:SP}).
Conversely, for times larger than  $+t_0$, the anti-resonance contribution vanishes, and the resonance contribution dominates. Therefore the time $t_0$ gives the time scale of  breaking of resonance-antiresonance symmetry.

To find  $-t_0$ we will assume that the imaginary part of the resonance energy is much smaller than its real part:
\begin{align} \label{ImEReE}
\left | \im E_{{\rm R}} \right| \ll \left| \re E_{{\rm R}} \right|.
\end{align}
In this case, and assuming $|E_{\rm R}|<2b$, we have $|\lambda_{\rm R}|\approx 1$, which is true for the parameters used in Fig.~\ref{fig:surv_res}. 

Under the assumption~\eqref{ImEReE}, we  set Eq.~\eqref{eq:rescontrib2} equal to zero to arrive at
\begin{align} \label{eq:t0}
 -t_0 = \frac{2 \log (\lambda_{\rm R}) - \pi i}{-i E_{\rm R}}.
\end{align}
This expression is approximately real because the energy $E_{\rm R}$ is approximately real and $\log( \lambda_{\rm R})$ is approximately pure imaginary.  Note that as long as \eqref{ImEReE} is satisfied, $t_0$ is independent of $ \im E_{{\rm R}}$.

It is interesting to compare $t_0$ with the ``Zeno time'' $t_Z$ discussed in Ref.~\cite{Barsegov}. This time is given by $t_Z = 1/E$, where $E$ is the unperturbed energy of the excited state. It is of the same order of magnitude as $t_0 \sim 1/E_R$  in Eq.~\eqref{eq:t0}.  The time $t_Z$ marks a transition from early non-exponential decay ($t\lesssim t_Z$) to exponential decay  ($t\gtrsim t_Z$).  It is called the  Zeno time due to the following connection with the  quantum Zeno effect~\cite{Zeno}:

The Zeno effect occurs when,  shortly after the excited state is prepared at $t=0$, an experimenter performs repeated measurements at times $\Delta t$, $2\Delta t$, $3\Delta t$, \textit{etc}., to test whether the state is still excited.  If $\Delta t$ is short enough, so that for $|t|<\Delta t$ the survival probability is parabolic,  then the repeated measurements ``freeze'' the survival probability and prevent it from decaying.  The time $t_Z$ is generally much larger than $\Delta t$, so that $t_Z$ gives an upper bound for $\Delta t$; it gives a time boundary beyond which there can be no Zeno effect.

Besides the relation $t_0\sim t_Z$, there is also the following connection between resonance-antriresonance symmetry and  the Zeno effect: the latter can only occur if the time derivative of the unmeasured survival probability vanishes at $t=0$. In terms of our analysis, this occurs due to the cancellation of resonance and antisonance terms that are linear in time. This cancellation is visualized in Fig.~\ref{fig:SP}(a), where the resonance and anti-resonance curves have opposite slopes around $t=0$ while they have about the same magnitude. The Zeno effect then is associated with the existence of an unbroken resonance-antiresonance symmetry. Once the symmetry is broken, the excited state can no longer be frozen by repeated measurements.

We have considered the case in which there is only one pair of resonance and anti-resonance. If there is more than one pair, each pair will have its own time  $t_0$. We could then consider the ratio of the sum of all the resonance contributions to the sum of all the antiresonance contributions, as a measure of the overall resonance-antiresonance symmetry. This symmetry would be broken after the largest time $t_0$, among all the pairs. 
 
We remark that when a resonance is  close to a band-edge or to an exceptional point, the non-exponential decay is enhanced \cite{Jittoh05, GGCalderon06, Garmon13,SGGO16}. For example, the enhancement due to the band edges (when $E_R\to\pm2$ or $\lam_R\to\pm 1$) can be seen in Eqs.~\eqref{eq:Alongt} and~\eqref{eq:fn6} as well as Eq.~(119) of Ref.~\cite{Hatano14}. Figure \ref{fig:Ratio}(c) shows that, when the resonance is close to the exceptional point, it never dominates over the anti-resonance; there is never a regime of exponential decay in the survival probability. 

In these cases, the anti-resonant component is no longer negligible as compared to the resonant component after the short time scale given by $t_Z=1/|E_R|$. The time  $t_Z$ then only gives an upper bound of validity of the short-time approximation in Eq.~\eqref{eq:rescontrib2}.  

\section{Extension to a spatially-continuous model}
\label{sec:FM}
\begin{figure}
\includegraphics[width=0.7\columnwidth]{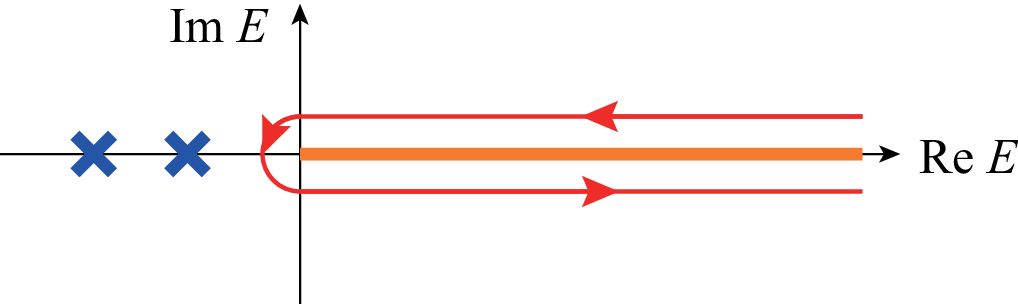}
\caption{The contour $C$ in Eq.~\eqref{eq:FM44}.
The branch cut of $\sqrt{E}$ lies on the positive real axis, on which also lies the scattering continuum.
Bound states can exist on the negative real axis.}
\label{fig9}
\end{figure}

\begin{figure}
\includegraphics[width=0.85\columnwidth]{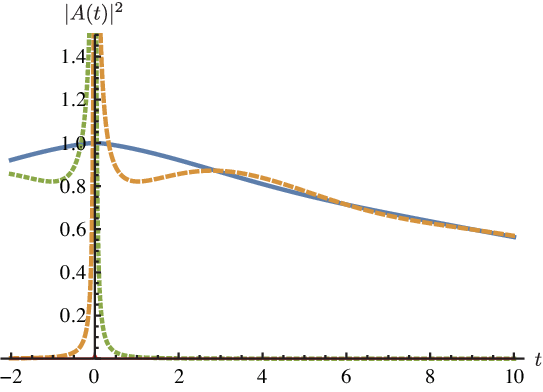}
\caption{\label{fig:survampFM} Survival probability $|A(t)|^2$  for the Friedrichs model (solid line); resonant-state contribution $\left|A_R(t)\right|^2$ (dashed line) and antiresonant-state contribution $\left|A_{AR}(t)\right|^2$ (dotted line), with the amplitudes $A_n(t)$ given in Eq.~\eqref{eq:FM6}. The bound-state contribution $\left|A_B(t)\right|^2$ can be barely seen as a small peak around $t=0$; it is an even function of time. The sum of resonant and anti-resonant contributions form an even function of time, but individually they break T-invariance. The parameters used are $\beta=0.5$ and $g=0.1$. For these parameters the resonance energy is $E_R=0.98 - 0.030 i$. The time scale for the breaking of resonance-antiresonance symmetry discussed in the text is $1/|E_R|= 1.02$}
\end{figure}

The results presented so far were based on a model with discrete space.  For models with continuous space, time-reversal symmetric expansions in terms of discrete eigenstates of the Hamiltonian  have been studied in Refs.~\cite{More71,GarciaCalderon76,Tolstikhin98,Tolstikhin06,Tolstikhin08,GarciaCalderon10,GarciaCalderon12,Brown16}. In these studies, the Hamiltonian has a scattering potential with compact support.   Here we will show that our results for the tight-binding models can be directly extended to models with a discrete state coupled to a continuous space,  such as the Friedrichs model \cite{Friedrichs48} discussed in Ref.~\cite{Lewenstein00}, with Hamiltonian
\begin{align} \label{eq:FM1}
 H &= \omega_1 |1\ket\bra 1| + \sum_k \omega_k |k\ket\bra k| \nonumber\\
 &+ \sum_k \left(V_k | 1\ket\bra k| + V_k^* | k\ket\bra 1| \right),
\end{align}
where $-\infty<k<\infty$, 
\begin{align} \label{eq:FM2}
 \omega_k = |k|,
\end{align}
and
\begin{align} \label{eq:FM3}
V_k = g  \sqrt{\frac{2\pi}{L}\frac{\sqrt{\beta\omega_k}}{\omega_k+\beta}}.
\end{align}
The survival amplitude $A(t)= \bra 1|e^{-i Ht}|1\ket $ may be expressed as
\begin{align} \label{eq:AFM}
A(t) = \sum_{n\in {\rm bound}} \bra 1|\phi_n\ket e^{-i E_n t}  \bra \phi_n|1\ket + A_{\rm cut}(t),
\end{align}
where the first term on the right-hand side is the contribution from the bound eigenstates of the Hamiltonian (with eigenvalues $E_{n}$) and the second term is the contribution from the scattering eigenstates, expressed as 
\begin{align} \label{eq:FM44}
A_{\rm cut}(t) =\frac{1}{2\pi i} \int_C \bra 1|\frac{1}{E-H}|1\ket e^{-i E t}  dE.
\end{align}
Here $C$ is the contour  shown in Fig.~\ref{fig9}; it surrounds the branch cut of the Green's function on the complex-energy plane.  The Green's function for real $E$ is given by (see Appendix~\ref{app:GFM})
\begin{align} \label{eq:FM4}
\bra 1|\frac{1}{E-H \pm i0}|1\ket = \left(E-\omega_1+ 2\pi g^2  \frac{\beta \pm i \sqrt{\beta E}}{\beta + E}\right)^{-1},
\end{align}
where we use the $+$ sign for the part of the contour $C$ just above the real axis and the $-$ sign for the part just below. We rearrange the Green's function as follows:
\begin{align} \label{eq:FM46}
&\bra 1|\frac{1}{E-H \pm i0}|1\ket 
\nonumber\\
&=\frac{\beta+E}{(E-\omega_1)(\beta+E)+ 2\pi g^2  (\beta \pm i \sqrt{\beta E})}\nonumber\\
&=\frac{(\beta+E)\left((E-\omega_1)(\beta+E)+ 2\pi g^2  (\beta \mp i \sqrt{\beta E})\right)}
{\left((E-\omega_1)(\beta+E)+ 2\pi g^2\beta\right)^2 +   (2\pi g^2)^2\beta E},
\end{align}
where in the last line we multiplied and divided the fraction by the complex conjugate of the denominator. Inserting this expression into Eq.~\eqref{eq:FM44} we note that the real part of the Green's function cancels when we integrate along the contour $C$; thus only the imaginary part remains and we obtain
\begin{align} \label{eq:FM45}
&A_{\rm cut}(t) \nonumber\\
&=\int_0^\infty   \frac{2g^2(E+\beta)\sqrt{\beta E} e^{-i E t} }{\left[(E+\beta)(E-\omega_1) + 2g^2\pi \beta\right]^2 +(2g^2\pi)^2\beta E} dE.
\end{align}
Even though the denominator is a quartic polynomial, one of its roots, $E=-\beta$, is not a pole because the numerator vanishes at this point. Therefore there are only three poles, corresponding to a bound-state energy $E_B$, a resonance $E_R$ and an anti-resonance $E_{AR}=E_R^*$. Using a partial-fraction expansion we have
\begin{align} \label{eq:FM5}
A_{\rm cut}(t) & =\int_0^\infty \sqrt{\beta E} e^{-iEt}\nonumber\\
&\times  \left(\frac{W_B}{E-E_B} + \frac{W_R}{E-E_R} + \frac{W_{AR}}{E-E_{AR}}\right) dE,
\end{align}
where $W_{AR}=W_R^*$.  The same expansion has been used in Ref.~\cite{Lewenstein00} to study the anti-Zeno effect. Equation~\eqref{eq:FM5} is analogous to the expansion~\eqref{eq:xrep}; we can write it as $A_{\rm cut}(t) =\sum_n A_n(t)$, where
\begin{align} \label{eq:FM6}
A_n(t) & =\int_0^\infty  \sqrt{\beta E} e^{-iEt} \frac{W_n}{E-E_n}  dE.
\end{align}
 Consequently, just as we did with the discrete lattice model,  we can decompose the survival amplitude into components that describe the self-generated breaking of resonance-antiresonance symmetry; see Fig.~\ref{fig:survampFM}.
 
  We can also estimate the time scale for the breaking  of resonance-antiresonace symmetry, as follows. The integral in Eq.~\eqref{eq:FM6} can be expressed in terms of the complementary error function~\cite{Lewenstein00}:
\begin{align} \label{eq:FM7}
A_n(t) & = W_n \sqrt{\beta }\left[\sqrt{\frac{\pi}{i t}} - \pi i \sqrt{E_n}e^{-i t E_n} {\rm erfc}\left(i\sqrt{i E_n t}\right)\right].
\end{align}
For $t<0$ and $-t \gg 1/|E_R|$ we can use an asymptotic expansion of the complementary error function to obtain for the resonance contribution 
\begin{align} \label{eq:FM8}
A_R(t) & = W_R \sqrt{\pi \beta E_n}\frac{-i}{2\left(i\sqrt{i E_R t}\right)^3} \left[1+ O\left(\frac{1}{E_R t}\right)\right],
\end{align}
which vanishes as $(E_Rt)^{-3/2}$ for $-t\gg1/|E_R|$. Note that this is {\it not}  the long-time inverse power decay (e.g., Eq.~\eqref{eq:fn9}), which occurs for positive times that are much larger than the lifetime of the resonant state. 

Similarly, the anti-resonant contribution vanishes for $t \gg 1/|E_R|$. Therefore $1/|E_R|$ gives the time scale for the breaking of resonance-antiresonance symmetry. As we discussed in Section~\ref{sec:tscale}, we may call this time scale the ``Zeno time'' because the Zeno effect occurs within this time scale.

\section{Concluding remarks}
\label{sec:conc}

In this paper we have described exponential decay in simple open quantum systems, using the T-symmetric expansion in Eqs.~\eqref{eq:xrep} and~\eqref{eq:FM5}. 
These expansions  include both resonance and antiresonance components in a symmetric way. Starting with these expansions, we did the following:

\begin{enumerate}
\item We showed that time-evolution breaks, by itself, the symmetry of the resonance and antiresonance components of a time-reversal invariant initial state. When this symmetry is broken, the time evolution closely matches the irreversible evolution of the isolated remaining component  (resonant-state component for $t>0$ or antiresonant-state component for $t<0$). 

\item  We introduced a parameter ($r$) that quantifies the degree of resonance-antiresonance symmetry breaking in the evolving quantum state. This symmetry is effectively broken when $|\log r|\gg1$.

\item We estimated the time that it takes for the breaking of resonance-antiresonance symmetry to occur, shortly after $t=0$. This time is approximately $t_Z\sim 1/E_R$, where $E_R$ is the real energy of the resonanant state; it coincides with  the Zeno time of Ref.~\cite{Barsegov}, during which the decay is non-exponential and during which the Zeno effect may occur if repeated measurements are done on the excited state.
 
\item We showed that for very long times, when the time evolution is non-exponential, the resonance-antiresonance symmetry is restored.
\end{enumerate}
 
If we associate the quantum-mechanical arrow of time with the resonance-antiresonance broken symmetry,   our results indicate that  the arrow of time appears spontaneously in quantum mechanics even if the initial condition is time-reversal invariant. For a decaying state, the arrow of time appears during the time range when the survival probability decays exponentially. Outside of this range the arrow of time disappears.

In general, the T-symmetric  expansion  is easier to compute than the expansion in terms of scattering states (Eqs.~\eqref{eqsurv} and~\eqref{eq:AljLam3}), because in the  former expansion, the scattering eigenstates are  decomposed into simpler terms. In some cases it may be advantageous to break T-symmetry by hand; this means deforming the integration contour as in Eq.~\eqref{eq:AljLam4} to explicitly  isolate the residue at complex poles of the Hamiltonian~\cite{Berggren82}. This would be a good approximation when the contributions from the  left-out poles may be neglected.  In general, however, the T-symmetric expansion has the advantage that it associates each pole with a specific contribution to the time-evolving state and allows us to see how each contribution evolves separately. Moreover, the T-symmetric expansion is free from unbounded, exponentially growing terms.

The formulation discussed here can be extended to more complex quantum dots and more leads, as described in Ref.~\cite{Hatano14}. We will have $N>2$ in Eq.~\eqref{eq:xrep}, in general, and there will be more complex-conjugate pairs of poles. Each pair will have its own time scale for its breaking of resonance-antiresonance symmetry. 

It would be interesting  to extend the present formulation to the Liouville equation for density matrices \cite{Hatano14}. For the latter,  we may be able to identify time scales for the emergence of the second law of thermodynamics, which presumably occurs as the symmetry of resonance-antiresonance pairs is broken by time evolution. The consideration of density matrices will also lead to the question of irreversibility in quantum-measurement processes.

\appendix

\section{A review of resonant and anti-resonant states}
\label{app:review-res}

We here present a brief review of the present knowledge about resonant and anti-resonant states.
See also Section~II of Ref.~\cite{Hatano14}.

For the moment, we consider the standard one-body one-dimensional Schr\"{o}dinger equation
\begin{align}\label{eqa10}
\left(-\frac{d^2}{dx^2}+V(x)\right)\psi(x)=E\psi(x),
\end{align}
since it is presumably more familiar to general readers. (We have put $\hbar^2/(2m)$ to unity for brevity.)
In Eq.~\eqref{eqa10}, we assume that $V(x)$ is a real potential with a compact support: $V(x)=0$ for $|x|>L$.
The dispersion relation for $|x|>L$ is given by
\begin{align}\label{eqa20}
E=k^2.
\end{align}

\begin{figure*}
\includegraphics[width=0.7\textwidth]{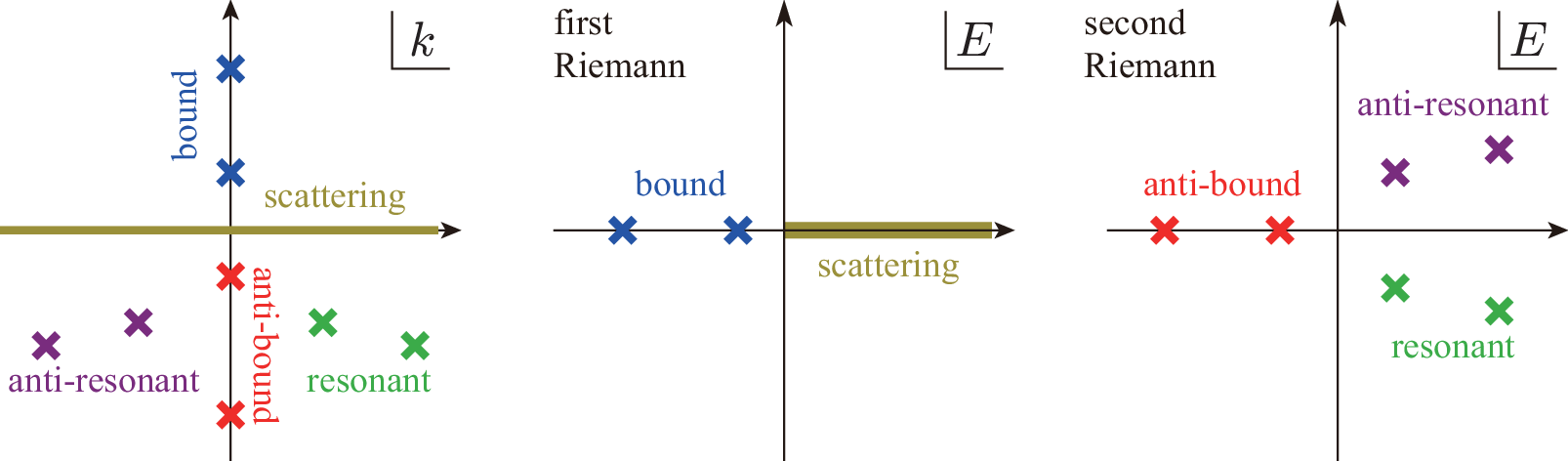}
\caption{A schematic view of all possible eigenstates of the Schr\"{o}dinger equation~\eqref{eqa10}.
See Table~\ref{atab1} for the precise attributes.}
\label{afig1}
\end{figure*}

The Schr\"{o}dinger equation~\eqref{eqa10} generally has the eigenvalues and the eigenfunctions listed in Table~\ref{atab1} and schematically shown in Fig.~\ref{afig1}.
\begin{table*}
\caption{All possible types of the eigenstates of the Schr\"{o}dinger equation~\eqref{eqa10}. By ``continuum" eigenvalue, we mean a continuous spectrum, while by ``discrete" eigenvalues, we mean point spectra. See Fig.~\ref{afig1} for a schematic view.}
\label{atab1}
\begin{tabular}{cccccc}
\hline
eigenfunction & eigenvalue & in the $k$ space && in the $E$ space & Riemann sheet \\
\hline\hline
scattering  state & continuum & real axis &&  positive real axis & first\\
\hline
bound  state & discrete & positive imaginary axis &\qquad& negative real axis & first\\
\hline
resonant state & discrete & the fourth quadrant && lower half plane & second\\
\hline
anti-resonant state & discrete & the third quadrant && upper half plane & second\\
\hline
anti-bound state & discrete & negative imaginary axis &\qquad& negative real axis & second\\
\hline
\end{tabular}
\end{table*}
The standard expansion of the unity and other quantities, such as Eq.~\eqref{eqsurv} (which is for the tight-binding model), uses all states in the first Riemann sheet but no states in the second.
In contrast, the new expansion given in Refs.~\cite{Hatano14,Tolstikhin98,GarciaCalderon10,GarciaCalderon12,Brown16}, such as Eq.~\eqref{eq:xrep} (which is again for the tight-binding model), uses all discrete states but no continuum states.

The scattering states are defined under the boundary condition
\begin{align}
\psi(x)=\begin{cases}
Ae^{ikx}+Be^{-ikx} &\quad\mbox{for $x<-L$},\\
Ce^{ikx}+De^{-ikx} &\quad\mbox{for $x>+L$},
\end{cases}
\end{align}
with the flux conservation $|A|^2-|B|^2=|C|^2-|D|^2$, because of which the wave number $k$ is real and the eigenvalue given by Eq.~\eqref{eqa20} is real positive.
All discrete states, on the other hand, are defined under the Siegert boundary condition~\cite{Gamow28,Siegert39,Peierls59,leCouteur60,Zeldovich60,Hokkyo65,Romo68,Berggren70,Gyarmati71,Landau77,Romo80,Berggren82,Berggren96,Madrid05,Hatano08,GarciaCalderon10}:
\begin{align}
\psi(x)\sim e^{ik|x|}
\quad\mbox{for $|x|>L$}.
\end{align}
This definition is indeed equivalent to the textbook definition of the resonant states as the poles of the S matrix~\cite{Hatano13}.
It includes a bound state as the case in which $k$ is a pure imaginary number with the positive imaginary part.

The other discrete states, namely the resonant, anti-resonant, and anti-bound states, all have negative imaginary part, and hence their eigenfunctions diverge in the limit $|x|\to\infty$. 
Because of this divergence, these states are often called unphysical and perhaps viewed skeptically, but it has been proved~\cite{Hatano08,Hatano09} that the resonant states  maintain the probability interpretation;
by incorporating the time-dependent part as in
\begin{align}
\Psi(x,t)=\psi(x)e^{-iEt},
\end{align}
we indeed realize that the spatial divergence is canceled out by the temporal decay due to the negative imaginary part of the energy eigenvalue.
From the point of view of the Landauer formula~\cite{Landauer57,Datta95}, the spatial divergence indicates the situation that the particle baths at the ends of the semi-infinite leads contain macroscopic numbers of the particles while the quantum scatterer in the middle contains only a microscopic number.

\section{Derivation of the expression~\eqref{eq:AljLam3} of the resonant-state component of the survival amplitude}
\label{app:A}
In this appendix we derive Eq.~\eqref{eq:AljLam3}.
The matrix elements of the eigenstates of the Hamiltonian appearing in Eq.~\eqref{eqsurv} are given by~\cite{Sasada11}
\begin{align}\label{dphik}
\bra d_1|\phi_{k\alpha}\ket &=  \bra d_1|\frac{1}{E_k-H+i0} H_1 |{k_\alpha}\ket \nonumber\\
&= -\sqrt{2} t_{2\alpha} \sin(k) \bra d_1|\frac{1}{E_k-H+i0} |d_2\ket
\end{align}
for $-\pi< k\leq \pi$,
where the matrix element of the Green's function is
\begin{align}\label{greenf}
& \bra d_1|\frac{1}{E_k-H+i0} |d_2\ket \nonumber\\
&=\frac{-g}{(E_k-\eps_{d_1})(E_k-\eps_{d_2}+e^{ik} \sum_{\alpha} t_{2\alpha}^2/b)-g^2}.
\end{align}
Equation~\eqref{dphik} is then expressed in terms of $\lambda=e^{ik}$ as
\begin{align}\label{dphilam}
\bra d_1|\phi_{k\alpha}\ket = \frac{-igt_{2\alpha} }{\sqrt{2}}  \frac{\lam-\frac{1}{\lam}}{f(\lam)}
\end{align}
with $f(\lam)$ given in Eq.~\eqref{flam}. 
Its absolute value squared is
\begin{align}\label{dphilam3}
\left|\bra d_1|\phi_{k\alpha}\ket\right|^2 &= \frac{g^2t_{2\alpha}^2}{2}  \left|\frac{\lam-\lam^{-1}}{f(\lam)}\right|^2 \nonumber\\
&=  \frac{g^2t_{2\alpha}^2}{2}  \frac{\left(\lam-\frac{1}{\lam}\right)^2}{f(\lam^{-1})-f(\lam)}\left(\frac{1}{f(\lam)}-\frac{1}{f(\lam^{-1})}\right)
\end{align}
because $\lambda^\ast=\lambda^{-1}$ for real $k$, and hence $f(\lambda)^\ast=f(\lambda^{-1})$.
In this expression, $f(\lam^{-1})-f(\lam)$ simplifies to 
\begin{align}\label{P4mP4}
f(\lam^{-1})-f(\lam) = -h(\lam) \sum_\alpha \frac{t_{2\alpha}^2}{b}\left(\lam-\frac{1}{\lam}\right),
\end{align}
where $h(\lam)=h(\lam^{-1})$ is given in Eq.~\eqref{hlam}. 
The survival amplitude in Eq.~\eqref{eqsurv} then takes the form
\begin{align} \label{eq:AljLam}
& A (t)  =  \sum_{n\in\mbox{\scriptsize bound}}\bra d_1|\phi_n\ket e^{-i E_nt}  \bra\phi_n|d_1\ket \nonumber\\
&- \frac{1}{2\pi i}
\int_O \frac{d\lambda}{\lam} \left(\lambda-\frac{1}{\lambda}\right)\exp\left[ib\left(\lambda+\frac{1}{\lambda}\right)t\right]
\nonumber\\
&\times
\frac{bg^2}{2 h(\lam)} \left[\frac{1}{f(\lam)}-\frac{1}{f(\lam^{-1})}\right],
\end{align}
where the integration contour $O$ is the counterclockwise unit circle on the $\lam$ complex plane.
Changing variables from $\lam$ to $\lam^{-1}$ for the term with $f(\lam^{-1})$, we have 
\begin{align} \label{eq:AljLam2}
& A (t)  =  \sum_{n\in\mbox{\scriptsize bound}}\bra d_1|\phi_n\ket e^{-i E_nt}  \bra\phi_n|d_1\ket \nonumber\\
&- \int_O \frac{d\lambda}{2\pi i \lam} \left(\lambda-\frac{1}{\lambda}\right)\exp\left[ib\left(\lambda+\frac{1}{\lambda}\right)t\right]
\frac{bg^2}{h(\lam)}\frac{1}{f(\lam)}.
\end{align}
 The function $f(\lam)$ in Eq.~\eqref{flam}  can be written as
\begin{align} \label{eq:P4f}
f(\lam) =\frac{b^2}{\lam^2} P_4(\lam),
\end{align}
where $P_4(\lam)$ is a fourth-order polynomial, 
\begin{align} \label{eq:P4fact}
P_4(\lam) =\prod_{n=1}^4 (\lam-\lam_n),
\end{align}
which has four roots $\lam_n$ corresponding to two bound-state eigenvalues and a  resonant-antiresonant pair of complex-conjugate eigenvalues of the Hamiltonian, for the parameters of Fig.~\ref{fig:SP}.

 The bound-state terms in Eq.~\eqref{eq:AljLam2} are residues of the integral over $\lam$ at the bound-state eigenvalues, so that they can be included in the integral over $\lam$ by adding counterclockwise integrations around these eigenvalues. This gives the final result in Eq.~\eqref{eq:AljLam3}.

\section{Proof of expression~\eqref{eq:xrep} of the survival amplitude}\label{app:Eq21}
In  Eq.~\eqref{eq:AljLam}, the function $h(\lam)$ in Eq.~\eqref{hlam} is proportional to a second-order polynomial as
\begin{align} \label{eq:AljLam22}
h(\lam)=-\frac{b}{\lam}(\lam-\lam_0)(\lam-\lam_0^{-1}),
\end{align}
where
\begin{align}
\lambda_0=\frac{1}{2}\left(-\frac{\varepsilon_1}{b}+\sqrt{\frac{{\varepsilon_1}^2}{b^2}-4}\right).
\end{align}
Using a partial-fraction expansion we have
\begin{align} \label{eq:AljLam23}
&\frac{1}{h(\lam)f(\lam)} = \frac{\lam^3}{b^3 (\lam-\lam_0)(\lam-\lam_0^{-1})\prod_{n=1}^4 (\lam-\lam_n)}\nonumber\\
&=\sum_{n=1}^4 \frac{W_n}{\lam-\lam_n} +   \frac{W_0}{\lam-\lam_0} + \frac{ W_{\bar 0}}{\lam-\lam_0^{-1}},
\end{align}
where the coefficients are given by 
\begin{align} \label{eq:AljLam24}
W_n &=  \lim_{\lam\to \lam_n} \frac{\lam-\lam_n}{h(\lam_n)f(\lam)},
\\ 
\label{eq:AljLam244}
W_0 &=   \lim_{\lam\to \lam_0} \frac{\lam-\lam_0}{h(\lam)f(\lam_0)} = \frac{\lam_0}{b} \frac{1}{\lam_0-\lam_0^{-1}} \frac{1}{f(\lam_0)},
\\
\label{eq:AljLam245}
W_{\bar 0} &=  \lim_{\lam\to \lam_0^{-1}} \frac{\lam-\lam_0^{-1}}{h(\lam)f(\lam_0^{-1})} = \frac{\lam_0^{-1}}{b}\frac{1}{\lam_0^{-1}-\lam_0} \frac{1}{f(\lam_0^{-1})} \nonumber\\
&= -\lam_0^{-2}W_0,
\end{align}
and in the last line we used
\begin{align} \label{eq:fl0}
f(\lam_0^{-1}) = f(\lam_0),
\end{align}
which follows from the fact that  when $h(\lam)=0$, \textit{i.e.}, when $\lam=\lam_0$ or $\lam=\lam_0^{-1}$,  we have $f(\lam)=f(\lam^{-1})=-g^2$ in Eq.~\eqref{flam}. 

In Eq.~\eqref{eq:AljLam} we can write
\begin{align} \label{eq:AljLam25}
\frac{1}{h(\lam)}\left[\frac{1}{f(\lam)}-\frac{1}{f(\lam^{-1})}\right] = \frac{1}{h(\lam)f(\lam)}- \frac{1}{h(\lam^{-1})f(\lam^{-1})}
\end{align}
since $h(\lam)=h(\lam^{-1})$. Introducing the partial-fraction expansion, Eq.~\eqref{eq:AljLam23}, into Eq.~\eqref{eq:AljLam25} we have
\begin{align} \label{eq:AljLam255}
&\frac{1}{h(\lam)f(\lam)}- \frac{1}{h(\lam^{-1})f(\lam^{-1})}
= \sum_{n=1}^4 \left(\frac{W_n}{\lam-\lam_n} -\frac{W_n}{\lam^{-1}-\lam_n} \right)\nonumber\\
&+ \frac{W_0}{\lam-\lam_0} +\frac{ W_{\bar 0}}{\lam-\lam_0^{-1}} 
- \frac{W_0}{\lam^{-1}-\lam_0} -  \frac{ W_{\bar 0}}{\lam^{-1}-\lam_0^{-1}}.
\end{align}
The second line in Eq.~\eqref{eq:AljLam255} can be shown to vanish identically
for $W_{\bar 0}=-\lam_0^{-2}W_0$. Hence, the terms involving $W_0$ and $W_{\bar 0}$ cancel, and Eq.~\eqref{eq:AljLam} reduces to 
\begin{align} \label{eq:AljLam27}
& A (t)  =  \sum_{n\in\mbox{\scriptsize bound}}\bra d_1|\phi_n\ket e^{-i E_nt}  \bra\phi_n|d_1\ket \nonumber\\
&- \frac{1}{2\pi i}
\int_O \frac{d\lambda}{\lam} \left(\lambda-\frac{1}{\lambda}\right)\exp\left[ib\left(\lambda+\frac{1}{\lambda}\right)t\right]
\nonumber\\
&\times
\frac{bg^2}{2} \left[ \sum_{n=1}^4 \frac{ W_n}{\lam-\lam_n}  -\sum_{n=1}^4 \frac{W_n}{\lam^{-1}-\lam_n}\right].
\end{align}
We remark  that the poles of $h(\lam)$ do not contribute to the integral.

Changing the integration variable from $\lam$ to $\lam^{-1}$ for the last term we get
\begin{align} \label{eq:AljLam28}
& A (t)  =  \sum_{n\in\mbox{\scriptsize bound}}\bra d_1|\phi_n\ket e^{-i E_nt}  \bra\phi_n|d_1\ket \nonumber\\
&- 
\int_O \frac{d\lambda}{2\pi i} \left(\lambda-\frac{1}{\lambda}\right)\exp\left[ib\left(\lambda+\frac{1}{\lambda}\right)t\right]
\nonumber\\
&\times
 \frac{bg^2}{\lam} \sum_{n=1}^4 \frac{W_n}{\lam-\lam_n}.
\end{align}
Expressing the bound-state terms as residues around the poles $\lam_n$ with $n\in\rm{bound}$ we obtain
\begin{align} \label{eq:AljLam29}
& A (t)  =   
\int_C \frac{d\lambda}{2\pi i} \left(\lambda-\frac{1}{\lambda}\right)\exp\left[ib\left(\lambda+\frac{1}{\lambda}\right)t\right]
\nonumber\\
&\times
 \frac{bg^2}{\lam}  \sum_{n=1}^4 \frac{W_n}{\lam-\lam_n},
\end{align}
where $C$ is the contour of Fig.~\ref{fig:Contour}. A further partial-fraction expansion gives
\begin{align} \label{eq:AljLam299}
 \frac{1}{\lam} \frac{W_n}{\lam-\lam_n} = \frac{W_n}{\lam_n} \left(\frac{1}{\lam-\lam_n}-\frac{1}{\lam}\right).
\end{align}
The isolated term $-1/\lam$ gives a vanishing integral because 
\begin{align} \label{eq:AljLam300}
&\int_C \frac{d\lambda}{2\pi i} \frac{1}{\lam} \left(\lambda-\frac{1}{\lambda}\right)\exp\left[ib\left(\lambda+\frac{1}{\lambda}\right)t\right]\nonumber\\
&=
\int_C \frac{d\lambda}{2\pi i} \frac{1}{ibt}  \frac{\partial}{\partial \lam} \exp\left[ib\left(\lambda+\frac{1}{\lambda}\right)t\right] = 0.
\end{align}
Therefore
\begin{align} \label{eq:AljLam301}
& A (t)  =   
\int_C \frac{d\lambda}{2\pi i} \left(\lambda-\frac{1}{\lambda}\right)\exp\left[ib\left(\lambda+\frac{1}{\lambda}\right)t\right]
\nonumber\\
&\times
 bg^2  \sum_{n=1}^4 \frac{W_n}{\lam_n}\frac{1}{\lam-\lam_n}.
\end{align}
For $t=0$, the residue of Eq.~\eqref{eq:AljLam301} at $\lam=\lam_n$ is the same as the residue of the Green's function at the pole $E_n$,
\begin{align} \label{eq:GFRes}
R_n \equiv \int_{C_n} \frac{dz}{2\pi i} \frac{1}{z-H} = {\rm Res}\left[\frac{1}{z-H}\right]_{z=E_n},
\end{align}
 where $C_n$ is a small contour surrounding the pole $E_n$ in a counterclockwise direction. In the following we will prove that $R_n = \Phi_n$, where $\Phi_n =|\phi_n\ket\bra{\tilde\phi}_n|$ is a dyad of discrete eigenstates of the Hamiltonian. 
 
 Since $H\Phi_n =\Phi_n H = E_n \Phi_n$, we can show by taking matrix elements that
\begin{align} \label{eq:GFRes2}
\bra j|\Phi_n|l\ket = \sum_{k\ne l}\frac{1}{E_n-\eps_j} \bra j|H_1|k\ket \bra k|\Phi_n|l\ket,
\end{align}
where $\eps_j =\bra j|H_0|j\ket$. On the other hand, the Green's function obeys the relation
\begin{align} \label{eq:GFRes3}
\frac{1}{z-H}= \frac{1}{z-H_0} + \frac{1}{z-H_0} H_1 \frac{1}{z-H},
\end{align}
 which leads to 
\begin{align} \label{eq:GFRes4}
\bra j|R_n|l\ket = \sum_{k\ne l}\frac{1}{E_n-\eps_j} \bra j|H_1|k\ket \bra k|R_n|l\ket.
\end{align}
Therefore $R_n$ and $\Phi_n$ obey exactly the same equation. Moreover, we have that $\Phi_n \Phi_{n'} = \delta_{nn'} \Phi_n$ and $R_n \Phi_{n'} = \delta_{nn'} \Phi_n$, which shows that $R_n$ and $\Phi_n$ have the same normalization. Therefore, we conclude that $R_n=\Phi_n$. 

By identifying the residue of Eq.~\eqref{eq:AljLam301} at $\lam=\lam_n$ with $R_n$ (and therefore with $\Phi_n$) for $t=0$,  we arrive to
\begin{align} \label{eq:AljLam30}
 -\left(\lambda_n-\frac{1}{\lambda_n}\right)  bg^2 \frac{W_n}{\lam_n}= \bra d_1|\phi_n\ket\bra{\tilde \phi}_n|d_1\ket,
\end{align}
which leads to 
\begin{align} \label{eq:AljLam31}
& A (t)  =   
\int_C \frac{d\lambda}{2\pi i} \left(\lambda-\frac{1}{\lambda}\right)\exp\left[ib\left(\lambda+\frac{1}{\lambda}\right)t\right]
\nonumber\\
&\times
 \sum_{n=1}^4 \frac{\bra d_1|\phi_n\ket\bra{\tilde \phi}_n|d_1\ket}{\lam_n^{-1}- \lam_n} \frac{1}{\lam-\lam_n}.
\end{align}
Finally, using Eq.~\eqref{eq:phipsi} we have
\begin{align} \label{eq:AljLam32}
 \frac{\bra d_1|\phi_n\ket\bra{\tilde \phi}_n|d_1\ket}{\lam_n^{-1}-\lam_n} 
 &=  \lam_n \frac{\bra d_1|\phi_n\ket\bra{\tilde \phi}_n|d_1\ket}{1-\lam_n^2} 
 \nonumber\\
 &= \lam_n \bra d_1|\psi_n\ket\bra{\tilde \psi}_n|d_1\ket,
\end{align}
which gives  Eq.~\eqref{eq:xrep}. 
\section{Derivation of the expression~\eqref{eq:Acut2} of the survival amplitude}
\label{app:B}
We will express the survival amplitude in terms of an integral of a Bessel function \cite{SGGO16}. The final result is given in Eq.~\eqref{eq:Acut2}.

We start by separating the survival amplitude in Eq.~\eqref{eq:xrep}  into two terms: one, $A_C(t)$, due to the clockwise contour around the unit circle in Fig.~\ref{fig:Contour}, and the other, $A_B(t)$, due to the small contours around the bound-state eigenvalues.  The survival amplitude is then  $A(t)=A_C(t)+A_B(t)$. 

Hereafter we consider the $A_C(t)$ term. Changing  variables from $\lambda$ to $k$ (with $ \lambda = e^{ik}$) we have
\begin{align} \label{eq:Aj}
& A_C(t)  =\frac{1}{2\pi i}
\sum_{n=1}^{2N} \int_{\pi}^{-\pi} dk (i e^{ik})(-2i \sin k) e^{2itb\cos k} 
\nonumber\\
&\times
\bra d_1|\psi_n\ket\frac{\lambda_n}{ e^{ik}-\lambda_n}\bra\tilde{\psi}_n|d_1\ket.
\end{align}
Multiplying and dividing the integrand by $(e^{-ik}-\lambda_n)$ and exchanging the integration limits we obtain
\begin{align} 
& A_C(t)  =\frac{-1}{2\pi i}
\sum_{n=1}^{2N} \int_{-\pi}^\pi dk (2 \sin k) e^{2itb\cos k} 
\nonumber\\
&\times
\bra d_1|\psi_n\ket\frac{b(1-e^{ik}\lambda_n)}{-2b\cos k  -E_n}\bra\tilde{\psi}_n|d_1\ket.
\end{align}
Hereafter we will assume that  $E_n$ is a resonant eigenvalue, and hence has a negative imaginary part. If it is a bound-state eigenvalue, for which the imaginary part is zero, we can add an infinitesimal imaginary part $-i \epsilon$ to $E_n$ and then take the limit $\epsilon \to 0$ at the end. If $E_n$  is an anti-resonant eigenvalue, which has a positive imaginary part, then the integration over $\tau$ in Eq.~\eqref{Itau} below should be done from $0$ to $-\infty$. 

Assuming $\im E_n<0$ we have
\begin{align} \label{Itau}
& A_C(t)  =\frac{-b}{2\pi i}
\sum_{n=1}^{2N} \int_{-\pi}^\pi dk (2 \sin k) e^{2itb\cos k} \nonumber\\
&\times
(-i ) \int_0^\infty d\tau e^{-i\tau(2b\cos k + E_n)}
 \bra d_1|\psi_n\ket(1-e^{ik}\lambda_n)\bra\tilde{\psi}_n|d_1\ket.
\end{align}
Terms that are even in $k$ in the second line of Eq.~\eqref{Itau} give a vanishing integral. Hence we have 
\begin{align} 
& A_C(t)  =
\sum_{n=1}^{2N} \frac{b\lambda_n}{\pi i} \int_0^\infty d\tau  e^{-i\tau E_n} \int_{-\pi}^\pi dk\, \sin^2 k\, e^{2i(t-\tau)b\cos k} \nonumber\\
&\times
 \bra d_1|\psi_n\ket\bra\tilde{\psi}_n|d_1\ket.
\end{align}
The integral over $k$ can be written in terms of the Bessel function $J_1$:
\begin{align} 
 \int_{-\pi}^\pi dk\, \sin^2 k\, e^{2i(t-\tau)b\cos k} = \pi \frac{J_1[2b(t-\tau)]}{b(t-\tau)}.
\end{align}
This gives
\begin{align} 
& A_C(t)  =
\sum_{n=1}^{2N} (-i)\lambda_n  \bra d_1|\psi_n\ket\bra\tilde{\psi}_n|d_1\ket  I(E_n,t),
\end{align}
where
\begin{align} 
  I(E_n,t)=  \int_0^\infty d\tau  e^{-i\tau E_n} \frac{J_1[2b(t-\tau)]}{t-\tau}.
\end{align}

Now we  change the integration variable $\tau$ to $t'=t-\tau$: 
\begin{align} 
 \label{Eq:B11}
 I(E_n,t) &=  \int_{-\infty}^t dt' e^{-iE_n(t-t')} \frac{J_1(2bt')}{t'} \nonumber\\
 &= e^{-iE_n t} \left(\int_{-\infty}^0 dt' +  \int_{0}^{t} dt' \right) e^{iE_n t'} \frac{J_1(2bt')}{t'}.
\end{align}
The integral from $-\infty$ to $0$ can be evaluated exactly in terms of the hypergeometric function ${}_2F_1$~\cite{web3}:
\begin{align} 
\label{HyperGF}
\int_{-\infty}^0  dt'& e^{iE_n t'} \frac{J_1(2bt')}{t'}  \nonumber\\
&=\int_{0}^\infty  dt' e^{-iE_n t'} \frac{J_1(2bt')}{t'}\nonumber\\
&=\frac{b}{ i E_n}\frac{\Gamma(1)}{\Gamma(2)} {}_2F_1  \left(\frac{1}{2},1; 2; \frac{4b^2}{E_n^2}\right) \nonumber\\
&=\frac{b}{ i E_n} \left(\frac{1}{2} + \frac{1}{2} \sqrt{1-\frac{4b^2}{E_n^2}}\right)^{-1}\nonumber\\
&= i \lambda_n^s,
\end{align}
where $s=1$ if  $|\lambda_n|<1$ (bound states) and $s=-1$ otherwise. Therefore we obtain
\begin{align} \label{eq:I9}
& I(E_n,t) = i e^{-iE_nt}\left[ \lambda_n^s 
 -i  \int_{0} ^t dt' \, e^{i E_n t'}\frac{J_1(2bt')}{t'} \right]
\end{align}
and
\begin{align} \label{eq:Acut22}
  &A_C(t) =
\sum_{n=1}^{2N}\lambda_n \bra d_1|\psi_n\ket \bra\tilde{\psi}_n| d_1 \ket \nonumber\\
&\times e^{-iE_nt}\left[ \lambda_n^s 
 -i  \int_{0} ^t dt' \, e^{i E_n t'}\frac{J_1(2bt')}{t'} \right].
\end{align}
Note that for the resonant state we do not need to include the contribution $A_B(t)$. We then obtain  Eq.~\eqref{eq:Acut2}.
\section{Derivation of the expression~\eqref{eq:Alongt} of the resonant-state component of the survival amplitude}
\label{app:C}
In this Appendix we  derive Eq.~\eqref{eq:Alongt} that gives the resonant-state component of the survival amplitude. 
We start with Eq.~\eqref{eq:Acut2}, which can be expressed as
\begin{align} \label{eq:Acut3}
  &A_{{\rm R}}(t)
=
 \bra d_1|\psi_R\ket \bra\tilde{\psi}_R| d_1 \ket \nonumber\\
&\times e^{-iE_Rt}\left[ 1
 -i  \lambda_R\left(\int_{0} ^\infty -\int_t^\infty\right) dt' \, e^{i E_R t'}\frac{J_1(2bt')}{t'} \right].
\end{align}
By analytic continuation of Eq.~\eqref{HyperGF}, the integral from $0$ to $\infty$ can be shown to be
\begin{align} \label{eq:Acut4}
 \int_{0} ^\infty dt' \, e^{i E_R t'}\frac {J_1(2bt')}{t'} =- i\lambda_R.
\end{align}
Therefore we have
\begin{align} \label{eq:Acut5}
  &A_{{\rm R}}(t)
=
 \bra d_1|\psi_R\ket \bra\tilde{\psi}_R| d_1 \ket \nonumber\\
&\times e^{-iE_Rt}\left[ 1-\lambda_R^2
 +i  \lambda_R \int_t^\infty dt' \, e^{i E_R t'}\frac{J_1(2bt')}{t'} \right].
\end{align}
For the states $|\phi_R\ket = \sqrt{1-\lambda_R^2} |\psi_R\ket$ and $\bra{\tilde \phi}_R| = \sqrt{1-\lambda_R^2} \bra{\tilde \psi}_R|$, Eq.~\eqref{eq:Alongt} follows. 

Up to this point we have considered positive $t$. In Appendix~\ref{app:LT} we will need to consider negative $t$. For a negative time $-t$, the first line of Eq.~\eqref{Eq:B11} gives 
\begin{align} 
 \label{Eq:B111}
 I(E_n,-t) &=   \int_{-\infty}^{-t} dt' e^{-iE_n(-t-t')} \frac{J_1(2bt')}{t'} \nonumber\\
 &=   \int_{t}^{\infty} dt' e^{iE_n(t-t')} \frac{J_1(2bt')}{t'} ,
\end{align}
where we changed integration variable from $t'$ to $-t'$ and used $J_1(-x)=-J_1(x)$. This gives
\begin{align} \label{eq:Acut55}
  &A_{{\rm R}}(-t)
=
  \bra d_1|\psi_R\ket \bra\tilde{\psi}_R| d_1 \ket \nonumber\\
&\times e^{iE_Rt} (- i  \lambda_R) \int_t^\infty dt' \, e^{-i E_R t'}\frac{J_1(2bt')}{t'} .
\end{align}
\section{Long-time approximation of the resonant-state component and estimation of $r(t)$ for large $t$}
\label{app:LT}
For large $t$ we neglect the purely exponential term in Eq.~\eqref{eq:Acut5} to obtain
\begin{align} \label{eq:LT1}
  A_{{\rm R}}(t)
=
  i  \lambda_R  \bra d_1|\psi_R\ket \bra\tilde{\psi}_R| d_1 \ket e^{-iE_Rt}
\int_t^\infty dt' \, e^{i E_R t'}\frac{J_1(2bt')}{t'} .
\end{align}
For $t'\gtrsim1$ the Bessel function is approximated as 
\begin{align} \label{eq:fn5}
   J_1(2bt') \approx \sqrt{\frac{1}{\pi bt'}} \sin\left(2bt'-\pi/4\right).
\end{align}
The integral in Eq.~\eqref{eq:LT1} can then be expressed in terms of the incomplete Gamma function,
\begin{align} \label{eq:fn6}
&  \int_t^\infty dt' \, e^{i E_R t'}\frac{J_1(2bt')}{t'} 
\approx   \frac{i \lam_R}{2i\sqrt\pi}  \nonumber\\
&\times\left[ e^{-i\pi/4} \left(\frac{ib}{2b+E_R}\right)^{-1/2} \Gamma\left(-\frac{1}{2}, -it(2b+E_R)\right)\right. \nonumber\\
& \left.-e^{i\pi/4} \left(\frac{ib}{-2b+E_R}\right)^{-1/2} \Gamma\left(-\frac{1}{2}, -it(-2b+E_R)\right)\right],
\end{align}
where
\begin{align} \label{eq:fn7}
\Gamma(a,z) = \int_z^\infty \tau^{a-1} e^{-\tau} d\tau.
\end{align}
For large $z$ the incomplete Gamma function is approximately given by
\begin{align} \label{eq:fn8}
\Gamma(a,z) \approx z^{a-1} e^{-z}.
\end{align}
Using this approximation for large $t\gg1/|E_R\pm 2b|$ in Eq.~\eqref{eq:fn6}, we obtain
\begin{align} \label{eq:fn9}
 &A_{{\rm R}}(t)
 =i  \lambda_R  \bra d_1|\psi_R\ket \bra\tilde{\psi}_R| d_1 \ket \frac{1}{2\sqrt\pi}  (bt)^{-3/2} 
\nonumber\\
&\times\left[ \left(\frac{b e^{-i\pi/4} }{2b+E_R}\right)  e^{2ibt} + \left(\frac{b e^{i\pi/4} }{2b-E_R}\right)  e^{-2ibt}\right],
\end{align}
which gives the power law decay of the survival probability in the form $t^{-3}$~\cite{Khalfin57, Garmon13, Hatano14,SGGO16}.

Similarly from Eq.~\eqref{eq:Acut55} we obtain for large negative time
\begin{eqnarray} \label{eq:fn10}
 A_{{\rm R}}(-t)
 &=&-i  \lambda_R  \bra d_1|\psi_R\ket \bra\tilde{\psi}_R| d_1 \ket \frac{1}{2\sqrt\pi}  (bt)^{-3/2} 
\nonumber\\
&\times&\left[ \left(\frac{be^{-i\pi/4} }{2b-E_R}\right)  e^{2ibt} + \left(\frac{be^{i\pi/4} }{2b+E_R}\right)  e^{-2ibt}\right].
\end{eqnarray}
For long times, the ratio of the resonance component of the survival amplitude to the anti-resonant component is
\begin{align} \label{eq:fn11}
 r(t)&=\left|\frac{A_R(t)}{A_R(-t)}\right|\nonumber\\
 &=\left|\frac{\left(\frac{e^{-i\pi/4} }{2b+E_R}\right)  e^{2ibt} + \left(\frac{e^{i\pi/4} }{2b-E_R}\right)  e^{-2ibt}}{
\left(\frac{e^{-i\pi/4} }{2b-E_R}\right)  e^{2ibt} + \left(\frac{e^{i\pi/4} }{2b+E_R}\right)  e^{-2ibt}}\right|.
\end{align}
If the imaginary part of $E_R$ is much smaller than its real part (e.g. the Fermi golden rule is applicable), then the ratio is close to $1$. 

\section{Derivation of the Green's function for the Friedrichs model}
\label{app:GFM}
In this appendix we obtain the Green's function in Eq.~\eqref{eq:FM4}. For the Friedrichs model, the Green's function is given by~\cite{Tasaki91}
\begin{align} \label{eq:GFM1}
G^\pm(E)\equiv \bra 1|\frac{1}{E-H \pm i0}|1\ket = \frac{1}{\eta^{\pm}(E)},
\end{align}
where 
\begin{align} \label{eq:GFM2}
\eta^\pm(E)&\equiv E-\omega_1-\sum_k \frac{V_k^2}{E^\pm-\omega_k}
\end{align}
and $E^\pm=E\pm i0$. In the continuous limit the summation approaches an integral as
\begin{align} \label{eq:GFM3}
\eta^\pm(E)&=E-\omega_1-g^2 \int_{-\infty}^{\infty} dk \frac{\sqrt{\beta\omega_k}}{\omega_k+\beta}
\frac{1}{E^\pm-\omega_k}\nonumber\\
&=E-\omega_1-2g^2 \int_{0}^{\infty} dk \frac{\sqrt{\beta k}}{k+\beta}
\frac{1}{E^\pm-k}.
\end{align}
In the second line we used $\omega_k=|k|$. We can integrate it over $k$ explicitly by changing the integration variable from $k$ to $u=\sqrt{k}$, which gives
\begin{align} \label{eq:GFM4}
&\int_{0}^{\infty} dk \frac{\sqrt{\beta k}}{k+\beta}
\frac{1}{E^\pm-k}\nonumber\\
&=-2 \sqrt{\beta}\int_{0}^{\infty} du \frac{u^2}{u^2+\beta}
\frac{1}{u^2 -E^\pm}\nonumber\\
&=-2 \frac{\sqrt{\beta}}{\beta+E^\pm} \int_{0}^{\infty} du \left(\frac{\beta}{u^2+\beta} + 
\frac{E^\pm}{u^2 -E^\pm}\right)\nonumber\\
&=\frac{-2 \sqrt{\beta}}{\beta+E^\pm} \left[ \sqrt{\beta}\arctan\frac{u}{\sqrt{\beta}}+\frac{\sqrt{E^\pm}}{2}  \ln\left(\frac{u+\sqrt{E^\pm}}{u-\sqrt{E^\pm}}\right)\right]_{0}^\infty\nonumber\\
&=\frac{-\pi}{\beta+E^\pm}\left(\beta + i\sqrt{\beta E^{\pm}}\right)=\frac{-\pi}{\beta+E^\pm}\left(\beta \pm i\sqrt{\beta E}\right).
\end{align}
In the last line we used the fact that the branch cut of the square root is along the positive $E$ axis. Inserting this result into Eq.~\eqref{eq:GFM3}, we obtain the Green's function in Eq.~\eqref{eq:FM4}.


\begin{acknowledgments}
We thank Savannah Garmon, Tomio Petrosky and Satoshi Tanaka for insightful discussions and  helpful suggestions. In particular, we thank S.~Garmon for critically reading the manuscript and suggesting various changes that helped improve clarity throughout the paper. GO acknowledges the Institute of Industrial Science at the University of Tokyo, the Department of Physical Science at Osaka Prefecture University, the Holcomb Awards Committee and the LAS Dean's office at Butler University for support of this work.
NH's research is partially supported by Kakenhi Grants No. 15K05200, No. 15K05207, and No. 26400409 from Japan Society for the Promotion of Science.
\end{acknowledgments}

\bibliography{hatano}

\end{document}